\documentclass[10pt,conference]{IEEEtran}
\pdfoutput=1
\usepackage{cite}
\usepackage{amsmath,amssymb,amsfonts}
\usepackage{algorithmic}
\usepackage{graphicx}
\usepackage{textcomp}
\usepackage{xcolor}
\usepackage[hyphens]{url}
\usepackage{fancyhdr}
\usepackage{hyperref}
\usepackage{lineno}

\usepackage{booktabs}

\makeatletter                   
\def\mdseries@tt{m}             
\makeatother                    
\usepackage[plain]{fancyref}
\usepackage[draft=true]{minted} 

\usepackage{color}
\usepackage{hyperref}           
\hypersetup{
    colorlinks=true,
    linkcolor=blue,
    filecolor=red,      
    urlcolor=magenta,
    breaklinks=true,            
}
\usepackage{breakurl}           

\usepackage{listings}
\usepackage{microtype}

\usepackage{multirow}
\usepackage{nth}
\usepackage{outlines}
\usepackage{pbox}
\usepackage{siunitx}
\usepackage{sourcecodepro}
\usepackage{tablefootnote}
\usepackage{xcolor-solarized}
\usepackage{xspace}
\usepackage{tabu}
\usepackage{tikz}
\usepackage{calc}
\usepackage[normalem]{ulem}
\usepackage{cleveref}

\usepackage{seqsplit}
\usepackage{pgfplots}
\pgfplotsset{compat=1.5}
\usetikzlibrary{chains,shadows.blur,fit,positioning}
\tikzset{fit margins/.style={/tikz/afit/.cd,#1,
    /tikz/.cd,
    inner xsep=\pgfkeysvalueof{/tikz/afit/left}+\pgfkeysvalueof{/tikz/afit/right},
    inner ysep=\pgfkeysvalueof{/tikz/afit/top}+\pgfkeysvalueof{/tikz/afit/bottom},
    xshift=-\pgfkeysvalueof{/tikz/afit/left}+\pgfkeysvalueof{/tikz/afit/right},
    yshift=-\pgfkeysvalueof{/tikz/afit/bottom}+\pgfkeysvalueof{/tikz/afit/top}},
    afit/.cd,left/.initial=2pt,right/.initial=2pt,bottom/.initial=2pt,top/.initial=2pt}
\tikzstyle{every picture}+=[remember picture]
\newcommand{\tikzmark}[1]{%
  \tikz[overlay,remember picture] \node (#1) at (0, 0.5ex) {};}
\makeatletter
\tikzset{
        from/.style args={#1 to #2}{
        above right={0cm of #1},
        /utils/exec=\pgfpointdiff
            {\tikz@scan@one@point\pgfutil@firstofone(#1)\relax}
            {\tikz@scan@one@point\pgfutil@firstofone(#2)\relax},
        minimum width/.expanded=\the\pgf@x,
        minimum height/.expanded=\the\pgf@y}}
\makeatother
\lstdefinestyle{revet}{
  language=C,
  basicstyle=\ttfamily\small\color{solarized-base02} ,
  numbers=left,
  numbersep=3pt,
  xleftmargin=1em,
  numberstyle=\tiny\color{solarized-base0},
  commentstyle=\color{solarized-green},
  morekeywords={pragma},
  keywordstyle=\bfseries\color{solarized-base03},
  morekeywords=[1]{int8,int16,int32,exit},
  morekeywords=[2]{load,store,fork,Foreach,foreach,replicate,replicate_bal,until,by,par,Reduce,Pipe},
  keywordstyle=[2]\bfseries\color{solarized-magenta},
  morekeywords=[3]{Reg,SRAM,DRAM,ReadView,WriteView,ModifyView,ReadIt,PeekReadIt,WriteIt,ManualWriteIt,ReadIterator,PeekReadIterator,WriteIterator,ManualWriteIterator},
  keywordstyle=[3]\bfseries\color{solarized-blue},
}
\lstdefinestyle{revetfootnote}{
  style=revet,
  basicstyle=\ttfamily\footnotesize\color{solarized-base02} ,
}
\lstdefinestyle{revetsmall}{
  style=revet,
  basicstyle=\ttfamily\scriptsize\color{solarized-base02} ,
}
\lstdefinestyle{revettiny}{
  style=revet,
  basicstyle=\ttfamily\tiny\color{solarized-base02} ,
}
\tikzset{>=latex}
\newcommand*\rfrac[2]{{}^{#1}\hspace{-.25ex}/\hspace{-.25ex}_{#2}}

\pgfplotscreateplotcyclelist{solarized}{%
  {draw=solarized-red,style=very thick,style=dotted,mark=*,mark size=1.00pt,mark options={solid,fill=solarized-red}},
  {draw=solarized-cyan,style=very thick,style=dashed,mark=*,mark size=1.00pt,mark options={solid,fill=solarized-cyan}},
  {draw=solarized-base03,style=very thick,mark=*,mark size=1.25pt},
}

\pgfplotscreateplotcyclelist{solarized-bars}{%
  {fill=solarized-red!50,draw=none},
  {fill=solarized-red,draw=none},
  {fill=solarized-cyan!50,draw=none},
  {fill=solarized-cyan,draw=none},
  {fill=solarized-magenta!50,draw=none},
  {fill=solarized-magenta,draw=none},
  {fill=solarized-green!50,draw=none},
  {fill=solarized-green,draw=none},
}
\pgfplotscreateplotcyclelist{solarized-single}{%
  {draw=solarized-base03,style=thick},
  {draw=solarized-base03,style=thick},
  {draw=solarized-base03,style=thick},
  {draw=solarized-base03,style=thick},
  {draw=solarized-base03,style=thick},
  {draw=solarized-base03,style=thick},
  {draw=solarized-base03,style=thick},
  {draw=solarized-base03,style=thick},
}

\newcommand{\name}{Revet\xspace}

\title{\name: A Language and Compiler for Dataflow Threads}

\makeatletter
\providecommand{\@LN}[2]{}
\makeatother

\newcommand{\hpcayear}{2024}

\newcommand{\hpcasubmissionnumber}{414}

\def\hpcacameraready{} 

\newcommand\hpcaauthors{Alexander C. Rucker$\dagger$, Shiv Sundram$\dagger$, Coleman Smith$\dagger$, Matthew Vilim$\ddagger$, Raghu Prabhakar$\ddagger$, \\ Fredrik Kj\o{}lstad$\dagger$, and Kunle Olukotun$\dagger$}
\newcommand\hpcaaffiliation{Stanford University$\dagger$, SambaNova Systems, Inc.$\ddagger$}
\newcommand\hpcaemail{acrucker@alumni.stanford.edu, shiv1@stanford.edu, csmith89@stanford.edu, matthew.vilim@sambanova.ai, \\ raghu.prabhakar@sambanova.ai, kjolstad@cs.stanford.edu, and kunle@stanford.edu}

\author{
  \ifdefined\hpcacameraready
    \IEEEauthorblockN{\hpcaauthors{}}
      \IEEEauthorblockA{
        \hpcaaffiliation{} \\
        \hpcaemail{}
      }
  \else
    \IEEEauthorblockN{\normalsize{HPCA \hpcayear{} Submission
      \textbf{\#\hpcasubmissionnumber{}}} \\
      \IEEEauthorblockA{
        Confidential Draft \\
        Do NOT Distribute!!
      }
    }
  \fi 
}

\fancypagestyle{camerareadyfirstpage}{%
  \fancyhead{}
  
  \fancyhead[C]{
    \ifdefined\aeopen
    \parbox[][12mm][t]{13.5cm}{\hpcayear{} IEEE International Symposium on High-Performance Computer Architecture (HPCA)}    
    \else
      \ifdefined\aereviewed
      \parbox[][12mm][t]{13.5cm}{\hpcayear{} IEEE International Symposium on High-Performance Computer Architecture (HPCA)}
      \else
      \ifdefined\aereproduced
      \parbox[][12mm][t]{13.5cm}{\hpcayear{} IEEE International Symposium on High-Performance Computer Architecture (HPCA)}
      \else
      \parbox[][0mm][t]{13.5cm}{\hpcayear{} IEEE International Symposium on High-Performance Computer Architecture (HPCA)}
    \fi 
    \fi 
    \fi 
    \ifdefined\aeopen 
      \includegraphics[width=12mm,height=12mm]{ae-badges/open-research-objects.pdf}
    \fi 
    \ifdefined\aereviewed
      \includegraphics[width=12mm,height=12mm]{ae-badges/research-objects-reviewed.pdf}
    \fi 
    \ifdefined\aereproduced
      \includegraphics[width=12mm,height=12mm]{ae-badges/results-reproduced.pdf}
    \fi
  }
  \fancyfoot[C]{}
}
\begin{document}
\maketitle

  \thispagestyle{plain}
  \pagestyle{plain}

\newcommand{\hpcaheight}{0mm}
\ifdefined\eaopen
\renewcommand{\hpcaheight}{12mm}
\fi


\begin{abstract}
  Spatial dataflow architectures such as reconfigurable dataflow accelerators (RDA) can provide much higher performance and efficiency than CPUs and GPUs.
  In particular, vectorized reconfigurable dataflow accelerators (vRDA) in recent literature represent a design point that enhances the efficiency of dataflow architectures with vectorization.
  Today, vRDAs can be exploited using either hard-coded kernels or MapReduce languages like Spatial, which cannot vectorize data-dependent control flow. 
  In contrast, CPUs and GPUs can be programmed using general-purpose threaded abstractions.

  The ideal combination would be the generality of a threaded programming model coupled with the efficient execution model of a vRDA. 
  We introduce \name: a programming model, compiler, and execution model that lets threaded applications run efficiently on vRDAs.
  The \name{} programming language uses threads to support a broader range of applications than prior parallel-patterns approaches, and our MLIR-based compiler lowers this language to a generic dataflow backend that operates on streaming tensors.
  Finally, we show that mapping threads to dataflow out-performs GPUs, the current state-of-the-art for threaded accelerators, by 3.8\texttimes.
\end{abstract}

\section{Introduction}
\label{sec:intro}
Spatial dataflow accelerators eliminate the overheads of modern von Neumann machines, including instruction fetch, dynamic scheduling, caching, and speculation, by statically scheduling computation.
In particular, vectorized reconfigurable dataflow accelerators (vRDA)~\cite{jouppi2017datacenter,jouppi2021ten,prabhakar2017plasticine,rucker2021capstan,prabhakar2021sambanova} are a new class of hardware that uses a large grid of compute and memory to exploit both \emph{vector} (SIMD) and \emph{pipeline} parallelism.
Furthermore, coarse-grained pipelining enables on-chip kernel fusion: a program can be pipelined across the entire chip without intermediate materializations to DRAM.
Overall, vRDAs maximize compute and memory bandwidth while lowering overhead.

However, vRDAs are limited by their programming model.
Currently, users program vRDAs with either hard-coded kernels directly expressed as low-level machine configurations~\cite{vilim2020gorgon,vilim2021aurochs} or hierarchical MapReduce code (e.g., the Spatial~\cite{koeplinger2018spatial} language).
Programming vRDAs with libraries of hard-coded kernels limits them to that small set of operations and prevents efficient fusion across operations, while prior MapReduce models limit programs to parallel loops over a control-flow-free inner loop body.
This eliminates algorithms with any data-dependent inner iteration because they require control flow, exceeding MapReduce's limits. 

Although vRDAs are more efficient, GPUs currently dominate the compute-accelerator market. 
Imperative programming models, including the \emph{threaded} SIMT model used by GPUs~\cite{buck2004brook}, are more powerful than MapReduce because they support parallelism over data-dependent structured control-flow like \lstinline!if! statements and \lstinline!while! loops.
This generality gap between MapReduce and SIMT inspires our key question: can we program vRDAs with thread-based programming languages?

In this paper, we introduce \name{}, a compiler that uses dataflow threading~\cite{vilim2021aurochs} to map a simple, yet expressive imperative language to vRDAs.
Dataflow threads increase the flexibility of vRDAs by moving control-flow decisions from a specialized control plane (with extremely limited bandwidth) to the faster data plane, which executes them as spatial routing decisions.
\name{} introduces a control-flow to dataflow lowering pass supporting a variety of control-flow  constructs including \lstinline!while! loops, \lstinline!if! statements, and nested parallel \lstinline!foreach! loops.
In turn, this control flow enables more asymptotically efficient algorithms to solve user-facing problems.

We describe several optimizations for \name.
These include efficient scratchpad orchestration for common access patterns, like data-dependent sequential reads and writes, with the ease of accessing caches.
Other optimizations lower the number of compute units needed for the resulting dataflow. 
Finally, we describe the composable abstract machine model that \name targets, which is based on the Aurochs vRDA~\cite{vilim2021aurochs}. 
By defining a machine model that uses a hierarchy of barriers to provide guarantees about control flow, \name{} makes it possible to compile arbitrary code that could not be compiled on Aurochs.
This includes code with nested \lstinline!while! loops, parallel-patterns \lstinline!foreach! loops inside dataflow threads, and dataflow threads inside \lstinline!foreach! loops---all with guaranteed correctness.

Our key contributions are:
\begin{enumerate}
  \item a vRDA abstract machine that adds hierarchical parallelism to Aurochs's per-lane control flow (\Cref{sec:dataflow}), 
  \item a language that captures parallelism and memory locality for applications with nested data-dependent control flow in a way that can map complex programs to vRDAs (\Cref{sec:lang}), and
  \item a compiler that optimizes and lowers our new language to streaming, vectorized, and pipelined dataflow on our abstract machine (\Cref{sec:implementation}).
\end{enumerate}

We demonstrate \name's flexibility by compiling a variety of applications drawn from data analytics, data structure traversal, geospatial analytics, and string analytics, none of which can be expressed in MapReduce.
We use cycle-accurate simulation to show that \name{} outperforms a V100 GPU by a geomean 3.8\texttimes{} on a 4.3\texttimes{} smaller vRDA, resulting in an area-adjusted speed up of 16\texttimes{}.
We also analyze how \name{} uses the vRDA's hardware units and demonstrate that our optimization passes enable significantly more efficient use of hardware resources.
\section{Background}
\name{} compiles to a machine model based on Aurochs~\cite{vilim2021aurochs}, a vRDA for dataflow threads. 
The current state of the art programming model for vRDA compilation is Spatial~\cite{koeplinger2018spatial}, which uses a parallel-patterns approach.
\name{} retains Spatial's support for explicit, user-facing parallelism while using dataflow threads to improve the flexibility of vectorization and add support for sequential control flow within parallel sections.

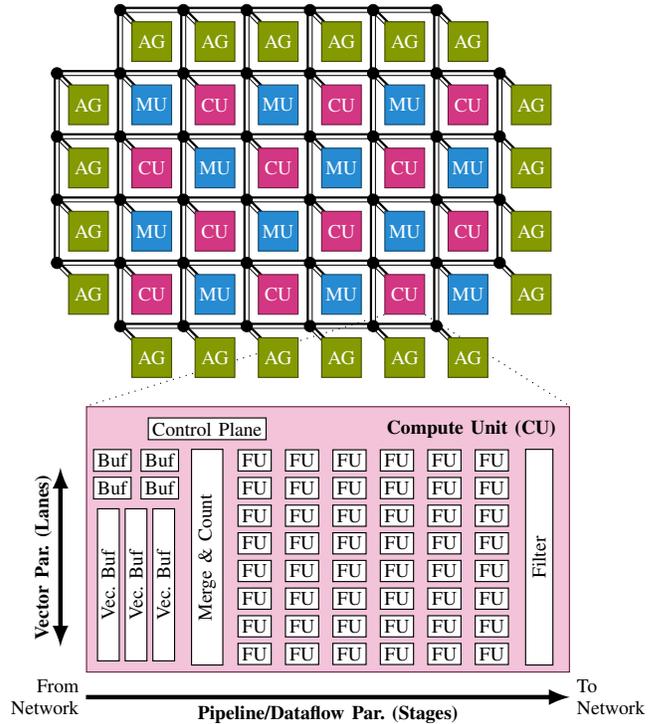
\begin{figure}
\color{black}
\centering

\pgfdeclarelayer{bg}    
\pgfsetlayers{bg,main}  
\resizebox{1.0\columnwidth}{!}{
\begin{tikzpicture}

\newcommand\gridx{7}
\newcommand\gridy{5}
\scriptsize
\foreach \i in {0,...,\gridx} {
  \foreach \j in {0,...,\gridy} {
    \newcommand\drawx{1}
    \newcommand\drawy{1}
    \ifthenelse{\(\i = 0 \OR \i = \gridx\) \AND \(\j = 0 \OR \j = \gridy\)}
    {}{
      \newcommand\nodetype{XXX}
      \newcommand\nodefill{white}
      \ifthenelse{\i = 0 \OR \i = \gridx \OR \j = 0 \OR \j = \gridy}{
        \renewcommand\nodetype{AG}
        \renewcommand\nodefill{solarized-green}
        \pgfmathtruncatemacro\xadd{\i+1}
        \pgfmathtruncatemacro\yadd{\j+1}
        \ifthenelse{\i = \gridx \OR \xadd = \gridx}{
          \renewcommand\drawx{0}
        }{}
        \ifthenelse{\j = \gridy \OR \yadd = \gridy}{
          \renewcommand\drawy{0}
        }{}
      }{
        \pgfmathtruncatemacro\sum{\i+\j}
        \ifthenelse{\isodd{\sum}}{
          \renewcommand\nodetype{CU}
          \renewcommand\nodefill{solarized-magenta}
        }{
          \renewcommand\nodetype{MU}
          \renewcommand\nodefill{solarized-blue}
        }
      }
      \pgfmathsetmacro\gridspace{0.8}
      \pgfmathsetmacro\posx{\gridspace*\i}
      \pgfmathsetmacro\posy{-\gridspace*\j}
      \pgfmathsetmacro\posxn{\gridspace*(\i+1)}
      \pgfmathsetmacro\posyn{-\gridspace*(\j+1)}
      \filldraw (\posx,\posy) circle[radius=2.0pt];
      \newcommand\nodeshift{4pt}
      \newcommand\drawshift{0.8pt}
      \newcommand\insep{1.5pt}
      \newcommand\nodedim{width("MU")+2*\insep}
      \node[fill=\nodefill,font=\color{white},rounded corners=0pt,minimum height=\nodedim,minimum width=\nodedim,inner sep=\insep,xshift=\nodeshift,yshift=-\nodeshift,draw=\nodefill!50!black,anchor=north west] at (\posx,\posy) (node-\i-\j) {\nodetype};
      \begin{pgfonlayer}{bg}
        \draw[very thin] ([yshift=-0.600*\drawshift,xshift=-0.600*\drawshift]\posx,\posy) -- ([yshift=-1pt-0.600*\drawshift,xshift=1pt-0.600*\drawshift]node-\i-\j.north west);
        \draw[thick] ([yshift=0.600*\drawshift,xshift=0.600*\drawshift]\posx,\posy) -- ([yshift=-1pt+0.600*\drawshift,xshift=1pt+0.600*\drawshift]node-\i-\j.north west);
      \end{pgfonlayer}
      \ifthenelse{\drawx = 1}{
        \draw[very thin] ([yshift=-\drawshift]\posx,\posy) -- ([yshift=-\drawshift]\posxn,\posy);
        \draw[thick] ([yshift=\drawshift]\posx,\posy) -- ([yshift=\drawshift]\posxn,\posy);
      }{}
      \ifthenelse{\drawy = 1}{
        \draw[very thin] ([xshift=\drawshift]\posx,\posy) -- ([xshift=\drawshift]\posx,\posyn);
        \draw[thick] ([xshift=-\drawshift]\posx,\posy) -- ([xshift=-\drawshift]\posx,\posyn);
      }{}
    }
  }
}
\newcommand\xoff{2.5}
\newcommand\yoff{-5.7}
\foreach \lane in {0,...,7} {
  \foreach \stage in {0,...,5} {
    \scriptsize
    \pgfmathsetmacro\posx{\xoff+0.6*\stage}
    \pgfmathsetmacro\posy{\yoff-0.35*\lane}
    \node[fill=white,minimum width={4pt},minimum height={3pt}, inner sep=1.5pt,draw] at (\posx,\posy) (fu-\stage-\lane) {FU};
  }
}
\newcommand\buffoff{1.1}
\foreach \ibs in {0,1} {
  \foreach \ib in {0,...,1} {
    \pgfmathsetmacro\posx{\xoff-\buffoff-0.1-0.6*\ibs}
    \pgfmathsetmacro\posy{\yoff-0.35*\ib}
    \node[fill=white,minimum width={4pt},minimum height={3pt}, inner sep=1.5pt,draw] at (\posx,\posy) (ib-scal-\ibs-\ib) {Buf};
  }
}
\foreach \ib in {-1,0,1} {
    \pgfmathsetmacro\posx{\xoff-\buffoff-0.4-0.35*\ib}
    \pgfmathsetmacro\posy{\yoff-4.5*0.35}
    \pgfmathsetmacro\minlen{0.35*5.5cm}
    \node[fill=white,rotate=90,minimum width={\minlen},minimum height={3pt}, inner sep=1.5pt,draw] at (\posx,\posy) (ib-vec-\ib) {Vec. Buf};
}
{
  \pgfmathsetmacro\posy{\yoff-0.35*3.5}
  \pgfmathsetmacro\minlen{0.35*7.8cm}
  {
    \pgfmathsetmacro\posx{\xoff+0.6*-1}
    \node[fill=white,minimum width={4pt},minimum height={3pt}, inner sep=1.5pt,draw] at (\posx,\yoff+0.4) (ctrl) {Control Plane};
    \node[fill=white,rotate=90,minimum width={\minlen},minimum height={3pt}, inner sep=2.5pt,draw] at (\posx,\posy) (mrg) {Merge \& Count};
  }{
    \pgfmathsetmacro\posx{\xoff+0.6*6}
    \node[fill=white,rotate=90,minimum width={\minlen},minimum height={3pt}, inner sep=2.5pt,draw] at (\posx,\posy) (filt) {Filter};
  }
}
\node[anchor=south east] (culabel) at (\xoff+3.9,\yoff+0.2) {\color{black}\bfseries Compute Unit (CU)};
\begin{pgfonlayer}{bg}
\node[fill=solarized-magenta!30!white,draw=solarized-magenta!50!black,fit=(ib-scal-1-0) (filt) (culabel)] (cu) {};
\end{pgfonlayer}

\begin{pgfonlayer}{bg}
  \draw[dotted] (node-5-4.south west) -- (cu.north west);
  \draw[dotted] (node-5-4.south east) -- (cu.north east);
\end{pgfonlayer}
{
  \pgfmathsetmacro\posy{\yoff-3.1}
  \node[align=right,anchor=north west,below left=0.00cm of cu] (in) {From\\ Network};
  \node[align=left,anchor=north east,below right=0.00cm of cu] (out) {To\\ Network};
  \draw[->,ultra thick] (in) -- (out) node[below, pos=0.5] {\bfseries Pipeline/Dataflow Par. (Stages)};
  \node[left=5.15cm of fu-5-0] (l0) {};
  \node[left=5.15cm of fu-5-7] (l7) {};
  \draw[<->,ultra thick] (l0) -- (l7) node[above, pos=0.5,rotate=90] {\bfseries Vector Par. (Lanes)};
}
  
\end{tikzpicture}
}
\\[-0.3\baselineskip]
\caption{A diagram showing \name{}'s layout and vector/pipeline parallelism across functional units (FUs) within a compute unit~\cite{vilim2021aurochs}. For simplicity, only eight lanes are shown.}
\label{fig:aurochs}
\end{figure}

\paragraph{Parallel-Patterns Dataflow}
Plasticine~\cite{prabhakar2017plasticine} is a vRDA that maps programs to a grid of compute and memory resources, as shown in \Cref{fig:aurochs}.
Specifically, Plasticine is a grid of vectorized compute units (CUs) and memory units (MUs), arranged in a checkerboard pattern and surrounded by DRAM address generators (AGs).
The CUs, MUs, and AGs are connected by a programmable network guaranteeing exactly-once, in-order delivery~\cite{zhang2019scalable}, and the entire chip runs at a fixed clock frequency.
Dataflow architectures are frequently network-limited~\cite{zhang2019scalable}, so efficiently using network resources is critical.
Plasticine's~\cite{prabhakar2017plasticine} network and per-unit input buffers are a mixture of vector (\SI{512}{b}) and scalar (\SI{32}{b}) resources.
To maintain 100\% dataflow throughput, each unit uses input buffers at the pipeline head to account for network path-length imbalances. 

SARA~\cite{zhang2021sara} lowers the Spatial~\cite{koeplinger2018spatial} language to streaming dataflow \emph{contexts,} which were each mapped to one or more CUs, MUs, or AGs based on their size.
First,  SARA maps the instructions inside the basic block to pipeline stages.
Then, SARA extracts control logic, which for parallel-patterns code is a nested series of counter-driven loops, and maps it to specialized control hardware.
Finally, SARA maps data values associated with innermost loops to vector dataflow while mapping other values to scalar dataflow.
SARA's controllers track one instance of control state per CU and can make one control-flow decision per cycle.
This control-plane/data-plane split results in a control plane with extremely limited bandwidth.

Furthermore, SARA's controllers are used to interpret data boundaries in on-chip links.
For example, if a link transmits three data elements $a,b,c$ it is up to the receiver to determine the correct allocation of these data to loop iterations. Thus, the transmitted data for $[a],[b,c]$ and $[a,b],[c]$ is identical.
For irregular programs, the complexity introduced by transmitting controller information can be significant.

\paragraph{Aurochs}
Aurochs~\cite{vilim2021aurochs} was introduced to support irregular database algorithms, like hash-joins and index traversal, which cannot be mapped to Plasticine's rigid parallel-patterns model.
Aurochs extends Plasticine's pipeline with new logical functionality~\cite{rucker2021capstan,vilim2020gorgon} while keeping a linear layout.
To support merge-sorting and inner joins, the input buffers can be interleaved using a merge unit and broadcast using a counter-based control unit.
After the pipeline-head logic has permuted the inputs, data then enters the pipeline, where six stages (each with an element-wise, statically-mapped instruction over 16 lanes) process it.
Finally, data exits through an optional filtering stage to the network.

Aurochs introduced the \emph{dataflow threads} model, where every \emph{thread} is simply a set of live values that are kept together in the pipeline.
Using the filtering stage and subsequent merging, Aurochs emulates basic control flow on these threads.
For example, filters can select between \lstinline!if! and \lstinline!else! branches, a data-dependent subset of threads for recirculation in a \lstinline!while! loop, or a set of threads to be dropped entirely.
When routing decisions send a thread's live values to a CU, the CU does its computation and yields a modified set of values representing the new thread state.

Aurochs has two key limitations that \name{} addresses. 
First, Aurochs can not use the lower-overhead scalar network because its dataflow threads lack the \emph{hierarchy} needed to associate a scalar with vectorized data.
This can lead to a shared value being copied into multiple threads and recirculated through the network repeatedly instead of being sent once and broadcast at the receiver.
In turn, this limitation prevents Aurochs from enabling fine-grained parallel patterns \emph{within} a data flow thread.

Second, Aurochs's ad hoc filtering mechanism lacks the ability to support arbitrary compiled control flow, because the machine model lacks an efficient mechanism for grouping and synchronizing threads.
Specifically, Aurochs used a timeout mechanism for synchronizing threads inside a recirculating region---if no activity is observed at the loop-head block for a given number of cycles, then Aurochs assumes that all threads have exited the loop body.
However, this mechanism breaks down for nested loops because a thread may be recirculating inside an inner loop for an arbitrary amount of time.

\section{A Generic Model of Dataflow}
\label{sec:dataflow}
Having introduced Aurochs~\cite{vilim2021aurochs}, the initial implementation of dataflow threads, we now discuss how \name{} formalizes and extends the dataflow threads abstraction, starting with the on-chip dataflow format.

\subsection{Structured-Link Tensor Format (SLTF)}
\label{sec:format}
Parallel-patterns dataflow operates on tensors coordinated using synchronized, counter-chain-driven controllers.
However, in \name{}, senders and receivers do not share synchronized controllers, so the sender must \emph{encode} its control decisions---and those made by upstream senders---so that they can be communicated to downstream units.
Encoding control-flow data can be done by selectively sending data to only some receivers or by changing metadata.
\name{} uses a structured-link tensor format (SLTF) to encode this control metadata.
The SLTF uses a small number of additional bits per on-chip link to count the number of elements being sent and encode a barrier indicating the number of nested loops that are being terminated.
Then, when a parallel-patterns reduction operation receives a loop termination, it sends the current value of the reduction value and resets the accumulator to its initial value.

\paragraph{Embedding Control with Data}
\name{} uses a structured on-chip data representation to capture live variables inside threads and hierarchy information across groups of threads.
The live variables within each thread are sent as parallel tensors, where ordering associates live values across tensors.
Hierarchy is encoded as done-tokens, or barriers ($\Omega_n$), to indicate the end of dimensions. For brevity, we represent barriers as $\Omega_n$ to indicate the end of dimension $n,$ starting with $\Omega_1$ to indicate the end of the lowest dimension.

Intuitively, the hierarchy metadata represents ragged $k$-dimensional tensors, where the number of dimensions is fixed but each dimension can have a variable size.
For example, the two-dimensional tensor [[0, 1], [2]] would be encoded as [0, 1, $\Omega_1,$ 2, $\Omega_2$] in the on-chip network.
Here, $\Omega_2$ implies an $\Omega_1,$ after element 2, due to the tensor dimensions forming a strict hierarchy and there being scalar elements in the tensor.

Adding this hierarchy to on-chip links using an out-of-band encoding is inexpensive.
We assume that at most one barrier can be sent per on-chip vector and that $n \leq 15.$
This is far lower than observed loop nesting levels and less than 1\% overhead relative to a 512-bit (16\texttimes32-bit) vector (assuming the vector link contains four bits for the barrier level and a length encoding that overlaps the last lane except for one bit).

\paragraph{Composability}
Handling the empty-tensor edge case is essential to composability: without precise control for empty tensors, reductions could not compose with downstream operations.
Therefore, to use embedded control metadata for synchronization, \name{} must precisely track empty lists.
For example, in our abstraction, the three 2-D tensors {[[]]} and {[[],[]]} and {[]} have unique representations ($\Omega_1,\Omega_2$ vs. $\Omega_1,\Omega_1,\Omega_2$ vs. $\Omega_2$).
Although all three of these tensors contain no actual data, they represent different control-flow structures: an outer loop running one iteration with a zero-length inner loop, an outer loop running twice with zero-length inner loops, or an outer loop that does not run.
Therefore, when passed to an additive reduction, they must yield distinct results: {[0]}, {[0,0]}, and {[]}.

\subsection{Streaming Tensor Primitives}
\label{sec:tensorprimitives}
Given a format for on-chip links that can embed and propagate control-flow decisions, we now describe the streaming primitives that implement local control decisions like parallelism and branching. 
These primitives are used for individual basic-block edges, and they respect our structured-link tensor format (\Cref{sec:format}), so they can be composed arbitrarily.
Together, they provide the sequencing, iteration, and selection needed for arbitrary algorithms. 
\name's machine model requires that primitives respect the SLTF for composability:
\begin{enumerate}
    \item Every barrier that enters a primitive exits that primitive exactly once, in order, and
    \item Thread data (stored in the SLTF) is not reordered with respect to barriers. Data can be reordered in between barriers.
\end{enumerate}
By obeying these conditions, \name's primitives can rely on the behavior of nested primitives to guarantee correctness.
For example, a \lstinline!while! loop containing an \lstinline!if! statement can rely on the \lstinline!if! statement not modifying barriers or reordering threads across barriers.
Similarly, an \lstinline!if! statement can contain a parallel-patterns \lstinline!foreach! loop on one of its branches---this is useful for cases like periodically loading a vector of data from DRAM to SRAM.

\begin{figure}
  \centering
\begin{minipage}[t]{0.5\columnwidth}
\centering
\small\textsf{Iteration Counts:}\\[.5\baselineskip]
{ \footnotesize
$t_1=2\quad t_2=3$ 

$t_3=1\quad t_4=3$
}
\\[.50\baselineskip]
\begin{tikzpicture}[auto, node distance=9mm, start chain=going below, 
  box/.style = {draw,rounded corners,blur shadow,fill=white, on chain,align=center}]
  \small
  \node[anchor=east] (scalar) at (-0.6, 0.15) {scalar};
  \node[anchor=east] (vector) at (-0.6,-0.15) {vector};
  \draw[->] (scalar) ++ (-1.0,0) -- (scalar);
  \draw[very thick,->] (vector) ++ (-1.0,0) -- (vector);

 \node[on chain] (b1)    {enter};      
 \node[box] (b2)    {fb-merge};      
 \node[box] (b3)    {filter};  
 \node[on chain] (b4)    {exit};     
 \begin{scope}[rounded corners,-latex]
  \draw (b3.-60) to   node[midway,right] {$D$}(b4.60);
  \draw (b1) -- (b2) node[midway] {$A$};
  \draw[very thick] (b2) -- (b3) node[midway] {$B$};
  \draw[very thick] (b3.230) -- ++(0,-0.3) -| node[midway,left] {$C$} ([xshift=-6mm]b2.west) |- 
   ([yshift=5mm]b2.130) -- (b2.130);
 \end{scope}
\end{tikzpicture}
\end{minipage}%
\begin{minipage}[t]{0.5\columnwidth}
\raggedright

\small\textsf{Tensor Abstraction:}\\[.5\baselineskip]
{
  \footnotesize
  \begin{tabular}{cl}
  $A:$ & $[t_1, t_2, t_3, t_4]$ \\
  $B:$ & $[[t_1, t_2, t_3, t_4],$\\
   & $[t_1, t_2, t_4], [t_2, t_4]]$\\
    $C:$ & $[[t_1, t_2, t_4], [t_2, t_4], []]$\\
  $D:$ & $[t_3, t_1, t_2, t_4]$ \\
  \end{tabular}
}
\\[1.0\baselineskip]
\small\textsf{SLTF:}\\[.5\baselineskip]
{ \footnotesize
  \begin{tabular}{cll}
  $A:$ & $t_1, t_2, t_3, t_4,\Omega_n$\\
  $B:$ & $t_1, t_2, t_3, t_4,\Omega_1$\\
       & $t_1, t_2, t_4,\Omega_1$\\ & $t_2, t_4,\Omega_1$\\
       & $\Omega_{n+1}$\\
  $C:$ & $t_1, t_2, t_4,\Omega_1$\\
       & $t_2, t_4,\Omega_1$\\ 
       & $\Omega_1$\\
  $D:$ & $t_3$ \\
   & $t_1$ \\
   & $t_2, t_4,\Omega_n$\\
  \end{tabular}
}
\end{minipage}
\lstset{style=revetfootnote}
\caption{
  The operation of a forward-backward merge unit (\lstinline!while! loop) showing how threads iterate repeatedly. This figure shows a scalar entry, under the assumption that each dataflow thread entering on link A will traverse links B and C multiple times.
}
\label{fig:fbmerge}
\end{figure}
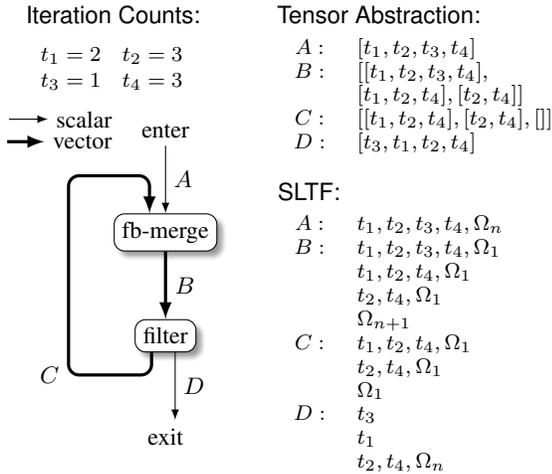

\paragraph{Element-Wise Operations}
Element-wise operations process one or more tensors: for example, two tensors may be added to yield a third tensor.
Memory operations are also element-wise operations: an allocation transforms a void value into a pointer, a read transforms an address into a result, and a write transforms an address and data into a void value.
Element-wise operations do not change the ordering, hierarchy, or number of dataflow threads.
Therefore, in this section's examples (\Cref{fig:mapreduce,fig:filtmerge,fig:fbmerge}), these operations can take place along any graph edge.

  \name{} provides memory ordering guarantees within a \emph{thread.}
  Therefore, the machine model must guarantee ordering for memory operations' side effects within a basic block.
To do so, it relies on data-free \emph{void} tokens like SARA's CMMC~\cite{zhang2021sara}: these are generated by memory operations as results and are inserted as operands.
Finally, these void tokens are carried through basic block transitions, like merges, to guarantee that basic blocks execute in order.

\paragraph{Expansion, Reduction, \& Flattening}
\begin{figure}
\centering
\begin{tikzpicture}[auto, node distance=9mm, start chain=going right, 
  box/.style = {draw,rounded corners,blur shadow,fill=white, on chain,align=center}]
\small

  \node[anchor=east] (scalar) at (1, 1.0) {scalar};
  \node[anchor=east] (vector) at (1, 0.7) {vector};
  \draw[->] (scalar) ++ (-1.0,0) -- (scalar);
  \draw[very thick,->] (vector) ++ (-1.0,0) -- (vector);

  \node[on chain] (b1)    {enter};      
 \node[box] (b2)    {counter};      
 \node[box] (b3)    {reduce};  
 \node[on chain] (b4)    {exit};     
 \begin{scope}[rounded corners,-latex]
  \draw (b1.40) to [bend left=20] node[midway,above] {$D$}(b4.130);
  \draw (b1) -- (b2) node[midway] {$A$};
  \draw[very thick] (b2) -- (b3) node[midway] {$B$};
  \draw (b3) -- (b4) node[midway] {$C$};
 \end{scope}
\end{tikzpicture}
\\[0.0\baselineskip]
\begin{minipage}[t]{0.45\columnwidth}
\raggedright

\small\textsf{Tensor Abstraction:}\\[.5\baselineskip]
{
  \footnotesize
  \begin{tabular}{rl}
  $A,C,D:$ & $[t_1, t_2]$ \\
  $B:$ & $[[t_{1.1}, t_{1.2}, t_{1.3}],$\\
       & $[t_{2.1}, t_{2.2}, t_{2.3}, t_{2.4}]]$\\
  \end{tabular}
}
\end{minipage}%
\begin{minipage}[t]{0.55\columnwidth}
\small\textsf{SLTF:}\\[.5\baselineskip]
{ \footnotesize
  \begin{tabular}{rll}
  $A,C,D:$ & $t_1, t_2, \Omega_n$ \\
  $B:$ & $t_{1.1}, t_{1.2}, t_{1.3}, \Omega_1$ \\
       & $t_{2.1}, t_{2.2}, t_{2.3}, t_{2.4},\Omega_{n+1}$\\
  \end{tabular}
}
\end{minipage}
\lstset{style=revetfootnote}
\caption{A \lstinline!foreach! loop: a 1-D tensor of threads is expanded into two dimensions and then contracted.
Hierarchical-tensor and streaming-barrier (SLTF) views of data are shown.
For simplicity, element-wise operations are elided. They could be added along any dataflow edge between complex primitives. }
\label{fig:mapreduce}
\end{figure}

Expansion primitives (\Cref{fig:mapreduce}) enlarge tensors to express map operations.
The simplest expansion primitive is broadcasting, which takes a $k$- and a $(k+1)$-dimensional tensor and repeats every element in the first tensor along the last dimension of the second one.
Counters  can also expand tensors: a counter takes three $k$-D tensors (min, max, and step) and transforms them into a $(k+1)$-D tensor.
Reduction (\Cref{fig:mapreduce}) uses an associative operation to coalesce the last tensor dimension into one element, lowering each barrier by one level.
Flattening also removes a level of hierarchy from barriers but leaves elements untouched.
Taken individually, these operations do not obey the SLTF constraints mentioned previously because they modify barrier levels.
However, an expansion/reduction pair can be used to wrap arbitrary code to implement a \lstinline!foreach! block.
Similarly, an expansion/flattening pair can be used to duplicate threads without adding a level of hierarchy, thus implementing a \lstinline!fork! statement.
Both \lstinline!foreach! and \lstinline!fork! obey the SLTF constraints.

\paragraph{Acyclic Subgraphs: Filtering \& Forward Merging}
Tensor filtering (\Cref{fig:filtmerge}) takes an element tensor and a predicate tensor and returns only the elements for which the predicate evaluates to true.
For example, an \lstinline!if! statement would use a filter operation to mask off elements so that each element goes to either the \lstinline!if! block or the \lstinline!else! block.
Barriers are passed through unmodified, creating two tensors from one moving forward through the pipeline.

Forward merging (\Cref{fig:filtmerge}) is used at the beginning of a basic block that has two \emph{forward} branches into it.
Merging interleaves elements from the lowest tensor dimension eagerly: whenever either input is ready to send, the merge can pass it through.
In \Cref{fig:filtmerge}, this is evident when $t_3,$ which branches onto a slow path, exits the merge last.
To preserve thread state, the merge can take multiple tensors (corresponding to all the live variables in a thread) and ensure that they are merged atomically.
Because the merge keeps per-thread data together, and threads within a hierarchy level are unordered, it preserves programming model correctness.

When the merge unit reaches a barrier in the streaming on-chip input, it stalls that link until it reaches an equal barrier in the opposite link.
This limits reordering (e.g., between two branches of the same \lstinline!if! statement) to one level of the tensor hierarchy, so threads in a parallel region do not cross barriers and remain synchronized to their parent thread.
By waiting for barriers to arrive, the \lstinline!if! statement is tolerant of network effects including delays and bandwidth limits.
\begin{figure}
\centering
\begin{tikzpicture}[auto, node distance=7.5mm, start chain=going right, 
  box/.style = {draw,rounded corners,blur shadow,fill=white, on chain,align=center}]
  \small
  \node[anchor=east] (scalar) at (1, 0.8) {scalar};
  \node[anchor=east] (vector) at (1, 0.5) {vector};
  \draw[->] (scalar) ++ (-1.0,0) -- (scalar);
  \draw[very thick,->] (vector) ++ (-1.0,0) -- (vector);

 \node[on chain] (b0)    {enter};
 \node[box] (b1)    {filter};
 \node[on chain] (b2)    {delay};      
 \node[box] (b4)    {fwd-merge};     
 \node[on chain] (b5)    {exit};     
 \begin{scope}[rounded corners,-latex]
  \draw[very thick] (b0) to node[midway,below] {$A$}(b1);
  \draw[very thick] (b1.40) to [bend left=20]  node[midway,above] {$C$}(b4.130);
  \draw (b1) to node[midway,below] {$B$}(b2);

  \draw (b2) -- (b4);
  \draw[very thick] (b4) -- (b5) node[midway] {$D$};
 \end{scope}
\end{tikzpicture}
\\[0.0\baselineskip]
\begin{minipage}[t]{0.5\columnwidth}
\raggedright

\small\textsf{Tensor Abstraction:}\\[.5\baselineskip]
{
  \footnotesize
  \begin{tabular}{rl}
  $A:$ & $[t_1, t_2, t_3, t_4, t_5]$ \\
  $B:$ & $[t_3]$ \\
  $C:$ & $[t_1, t_2, t_4, t_5]$ \\
  $D:$ & $[t_1, t_2, t_4, t_5, t_3]$ \\
  \end{tabular}
}
\end{minipage}%
\begin{minipage}[t]{0.5\columnwidth}
\small\textsf{{SLTF}:}\\[.5\baselineskip]
{ \footnotesize
  \begin{tabular}{rll}
  $A:$ & $t_1, t_2, t_3, t_4, t_5,\Omega_n$\\
  $B:$ & $t_3,\Omega_n$\\
  $C:$ & $t_1, t_2, t_4, t_5,\Omega_n$\\
  $D:$ & $t_1, t_2, t_4, t_5, t_3,\Omega_n$\\
  \end{tabular}
}
\end{minipage}
\lstset{style=revetfootnote}
\caption{
  In a filter-merge operation (\lstinline!if! statement), a vector of threads is partitioned into two vectors, one for each branch. 
  Here, link B is mapped as scalar to avoid overprovisioning network resources for a rare execution case. If links B and C were equally common, both could be mapped to vector dataflow resources at the cost of additional network congestion.
}
\label{fig:filtmerge}
\end{figure}

\paragraph{Cyclic Subgraphs: Forward-Backward Merging}
Like forward merging, forward-backward merging (\Cref{fig:fbmerge}) interleaves incoming threads.
However, unlike forward merging, forward-backward merging combines tensors resulting from backward branches (e.g., at the head of a \lstinline!while! loop).
Forward merging would not work for this case: the backward branch can only send a barrier \emph{after} the merge sends a barrier, but the merge can only send a final barrier once it receives one from the backward branch.
Therefore, forward-backward merge uses different logic to break this would-be cyclic dependency.
Intuitively, the forward-backward merge at the loop header takes a 1-D tensor of input elements at a time and iterates it to form a 2-D tensor of executed loop bodies.

A natural loop will have one header block, which is the meeting point of all forward edges into the loop and all backward edges, and the forward-backward merge is located at this loop header.
The forward-backward merge primitive starts by outputting values from the forward branch into the loop body until it receives a done-token.
The done-token causes the merge to stall inputs on the forward branch, and the loop header will use barrier semantics to ensure the loop body is empty before allowing more threads to enter.
Because the loop header is the sole entry point to a natural loop, it can \emph{reassign} barrier levels inside its loop as long as barriers exiting the loop are correct.
Specifically, the loop header \emph{adds} a level to incoming barriers, so it can reserve the lowest barrier $\Omega_1$ for checking whether the loop body is empty.
The merge will send a $\Omega_1$ token to terminate its sent data; it will continue to send this token every time it appears at the backward-branch input.

When the loop body is empty (all the threads executing the while loop have terminated), the backward branch will receive two $\Omega_1$ tokens in a row, which will cause the forward-backward merge to send a done token at one level higher than that originally received on the forward-branch link.
Edges leaving the body then lower all barriers by one level, eliminating the added $\Omega_1$ barriers and restoring input barriers to their correct levels.
This ensures that cyclic regions respect the same barrier constraints as acyclic ones, making them composable.
Unlike Aurochs, the lack of timeouts for detecting loop completion makes this abstraction usable for loops with arbitrarily long loop bodies (like nested while loops).

\subsection{Mapping to Virtual Hardware}
Finally, these primitives must be mapped to a hardware model.
We assume an overall compute-unit structure based on Aurochs and Plasticine~\cite{vilim2021aurochs,prabhakar2017plasticine}. 
In our model, mergers, counters, and broadcasts are at the beginning of the pipeline. 
The pipeline-head logic will stall inputs as needed to meet merging constraints, wait for all inputs to be available for element-wise operations, and send the correct barriers through the pipeline.
For example, the pipeline-head logic for a forward merge will take elements from each input and concatenate them until a barrier appears on one input.
Once the barrier arrives at an input, no further elements will be taken from it until an equivalent barrier appears on the other input; at that point, both barriers will be dequeued and sent as a single barrier.
Broadcasts are handled by repeating the element at the head of an input buffer across all valid lanes, and popping from the input once the corresponding barrier ($\Omega_1$ for a one-level broadcast, $\Omega_2$ for two-level, etc.) arrives.
Element-wise operations happen inside the pipeline, with the barriers inserted by the pipeline-head logic propagated unmodified.
Reductions, filters, flattening, and vector-to-scalar conversion happen at the end of the pipeline, along with any reduction in barrier level.

As shown in \Cref{fig:fbmerge}, on-chip SLTF links can be allocated to scalar or vector resources.
Because they contain their own metadata, SLTF links are agnostic to the bandwidth of network resources---a scalar link can send up to one data element and one barrier per cycle, while a vector link can send up to 16 data elements and one barrier per cycle.
For example, a vector with two elements and one barrier $(t_1,t_2,\Omega_1)$ can be sent on a vector link in one cycle, but would require two cycles on a scalar link: $t_1,$ then $t_2, \Omega_1.$
If two barriers are sent $(\Omega_1, \Omega_2),$ two cycles would be required on both vector and scalar links.
Because \name{} primitives do not consider timing information, sending a vector on a scalar link over multiple cycles is semantically equivalent to sending it all at once on a vector link.

SLTF links are allocated to network resources based on expected bandwidth.
For example, a \lstinline!while! loop with a high average trip count would use a scalar entry and vector backedge link to save resources at the loop header.
Conversely, a \lstinline!while! loop written to scan an open-addressed hash table may be expected to rarely loop, and may be provisioned with a vector entry and scalar backedge.

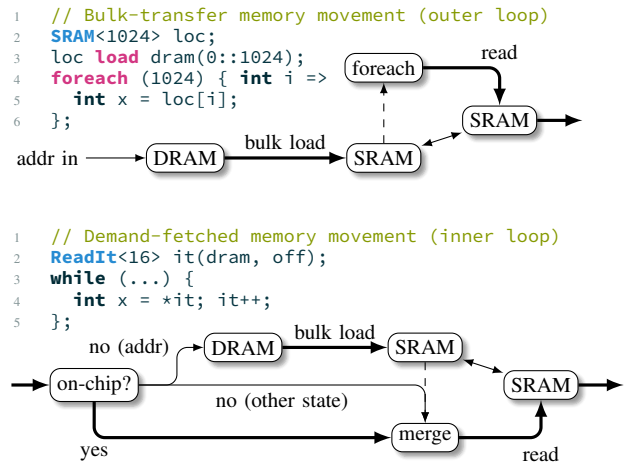
\begin{figure}
  \begin{lstlisting}[style=revetsmall]
  // Bulk-transfer memory movement (outer loop)
  SRAM<1024> loc;
  loc load dram(0::1024);
  foreach (1024) { int i =>
    int x = loc[i];
  };
  \end{lstlisting}
  \vskip-3.5\baselineskip
  \begin{tikzpicture}[auto, node distance=5mm, start chain=going right, x=2.2cm,y=0.5cm,
  box/.style = {draw,rounded corners,blur shadow,fill=white, align=center}]
    \footnotesize
  \node[box] (b2) at (0.8,1) {DRAM};      
  \node[box] (b3) at (2,1)    {SRAM};  
  \node[box] (b4) at (2.7,2)    {SRAM};     
  \node[box] (b5) at (2,3.4)    {foreach};     
  \node[fit=(b3) (b4)] (pmu) {};
    \node[above left] at (pmu.south east) {}; 
 \begin{scope}[rounded corners,-latex]
   \draw[very thick] (b2) -- (b3) node[midway,above] {bulk load};
   \draw[very thick] (b5) -| (b4) node[midway,above] {read};
   \draw (b2) ++(-0.6,0) -- (b2) node[pos=0,left] {addr in};
   \draw[dashed] (b3) -- (b5);
   \draw[very thick] (b4) -- ++ (0.5,0);
   \draw[<->] (b3) -- (b4);
 \end{scope}
\end{tikzpicture}
\\[-0.2\baselineskip]
  \begin{lstlisting}[style=revetsmall]
  // Demand-fetched memory movement (inner loop)
  ReadIt<16> it(dram, off);
  while (...) {
    int x = *it; it++;
  };
  \end{lstlisting}
  \vskip-1.0\baselineskip
  \begin{tikzpicture}[auto, node distance=5mm, start chain=going right, x=2.2cm,y=0.5cm,
  box/.style = {draw,rounded corners,blur shadow,fill=white, align=center}]
    \footnotesize
  \node[box] (b1) at (0,0) {on-chip?};      
  \node[box] (b2) at (0.9,1) {DRAM};      
  \node[box] (b3) at (2,1)    {SRAM};  
  \node[box] (b4) at (2.7,0)    {SRAM};     
  \node[box] (b5) at (2,-1.4)    {merge};     
  \node[fit=(b3) (b4)] (pmu) {};
    \node[below left] at (pmu.north east) {}; 
 \begin{scope}[rounded corners,-latex]
   \draw[very thick] (b1) |- (b5) node[pos=0.50,below] {yes};
   \draw (b1) -| (b5) node[pos=0.25,below] {no (other state)};
   \draw[very thick] (b2) -- (b3) node[midway,above] {bulk load};
   \draw[very thick] (b5) -| (b4) node[midway,below] {read};
   \draw (b1) -- ++(0.5,0) |- (b2) node[midway,left] {no (addr)};
   \draw[dashed] (b3) -- (b5);
   \draw[very thick] (b1) ++ (-0.5,0) -- (b1);
   \draw[very thick] (b4) -- ++ (0.5,0);
   \draw[<->] (b3) -- (b4);
 \end{scope}
\end{tikzpicture}
\caption{
  Above, Spatial~\cite{koeplinger2018spatial} requires that memory is explicitly transferred before the start of a parallel section.
  Below, \name{} uses control flow to coordinate transfers without stalls.
}
\label{fig:hitmiss}
\end{figure}

\section{The \name{} Language}
\label{sec:lang}
In the previous section, we formalized the dataflow threads machine model.
\name{} compiles a structured and imperative programming language, which is familiar to many programmers, to this abstract machine.

\subsection{Key \name{} Language Features}
The language has two new features compared to languages like C.
First, \name{} requires user-annotated parallelism in the form of \lstinline{foreach}, \lstinline{replicate}, and \lstinline{fork} statements.
Inside parallel regions, \name{} supports the fine-grained control flow expected in an imperative language.  
Second, \name{} uses iterators to efficiently orchestrate DRAM to SRAM transfers for data-dependent access patterns inside these sequential sections.

\paragraph{Flexible Threaded Parallelism}

By default, \name{}'s code execution is sequential, with mutable variables.
To enable parallelism, \name{} has explicitly parallel \lstinline{foreach} loops to eliminate any potential barriers to parallelism (e.g., aliasing); these \emph{child} loop bodies correspond to \emph{threads}.
Each thread's program statements run sequentially, but the execution order across threads is unsequenced.
Threads inside a \lstinline{foreach} have a read-only view of their parent's variables, but they can dereference pointers allocated by the parent to perform memory writes.
Finally, a \lstinline{foreach} thread may return a value, which is associatively reduced and returned to the parent.

\name{}, unlike prior work (such as Spatial~\cite{koeplinger2018spatial}, Plasticine~\cite{prabhakar2017plasticine}, or GPUs), 
supports flexible nested parallelism, where lanes of a vector-parallel program can have further-nested vector loops.
Nested parallelism enables scalar to vector \emph{broadcasting}, which uses fewer on-chip resources, and flexible parallelism allows vector regions to be nested inside other vector regions.
This is important for emulating caches: all threads start in a vector outer region and either traverse a vector cache hit path or a scalar miss path with a further-nested vector DRAM load.
Without this flexibility, the vector hit path would preclude vectorization of DRAM loads.

\name{} uses a two-step approach to extract parallelism beyond a single vectorized datapath.
First, a \lstinline{foreach} loop transitions code execution from scalar to vector.
Next, a \lstinline{replicate} statement transitions from a \emph{single vector} dataflow to \emph{multiple scalar} dataflow.
Finally, a \lstinline{foreach} or \lstinline{while} loop can be used inside the \lstinline{replicate} to finally create multiple vector dataflow pipelines.
To build \lstinline{replicate}, \name{} uses the filter operator to distribute threads across the parallel inner regions and the forward-merge operator to combine them at the end.
\begin{figure}
  \centering
  \tikzstyle{omegas}=[font=\scriptsize\color{red},inner sep=2pt]
\begin{tikzpicture}[auto, node distance=9mm, x=2.5cm, y=0.85cm,
  box/.style = {draw,rounded corners,blur shadow,fill=white, align=center}]
  \footnotesize
  \foreach\x in {3,2,1,0}
    \node[fill=solarized-base3,draw,dotted,anchor=north west,minimum width=190,minimum height=55] at (-0.4+\x*0.05,-3.80-\x*0.13) (replicate\x) {};
  \draw[dotted] (-0.60,-1.3) -| (2.7,-8.8) -| cycle;
  \draw[dotted] (-0.45,-2.5) -| (2.6,-7.6) -| cycle;
  \coordinate (in) at (0.5,-1.3);
  \coordinate (in2) at (0.5,-0.9);
  \draw[->] (in2) -- (in);
  \node[anchor=north east,omegas] at (in2) {$\Omega_0$};
  \node[anchor=east,omegas] at (0.5,-1.8) {$\Omega_1$};
  \node[box,draw=solarized-blue,text=solarized-blue] at (1.00,-1.8) (fetchA) {fetch (line 7)};
  \node[box,draw=solarized-blue,text=solarized-blue] at (2.00,-1.8) (storeA) {store (line 7)};
  \node[box,draw=solarized-base03,text=solarized-base03] at (1,-3.1)   (init) {init (lines 14--15)};
  \node[box,draw=solarized-base03,text=solarized-base03] at (1,-4.5) (blkA) {(line 22)};
  \node[box,draw=solarized-base03,text=solarized-base03] at (1,-5.5) (blkB) {(lines 22--24)};
  \node[box,draw=solarized-blue,text=solarized-blue] at (0,-4.5) (fetchB) {fetch (line 21)};
  \node[box,draw=solarized-blue,text=solarized-blue] at (0,-5.5) (storeB) {store (line 21)};
  \node[box,draw=solarized-base03,text=solarized-base03] at (1,-7)   (flush) {flush (line 27)};
  \node[box,draw=solarized-blue,text=solarized-blue] at (0.0,-8.25) (fetchC) {fetch (line  9)};
  \node[box,draw=solarized-blue,text=solarized-blue] at (1.0,-8.25) (storeC) {store (line  9)};

  \draw[very thick,->] (blkA) -- (blkB) node[midway,right,align=left,anchor=west] {hit};
  \draw[very thick,->] (blkA) -- (blkB) node[midway,left,anchor=east,omegas] {$\Omega_3$};
  \draw[very thick,->] (blkB) -- ++ (0.4,0) |- (blkA);
  \node (be-turn) at (1.4,-5.5) {};

  \draw[->] (blkA) -- (fetchB) node[midway,right,align=center,anchor=south] {miss};
  \draw[very thick,->] (fetchB) -- (storeB) node[midway,left,omegas] {$\Omega_4$};
  \draw[->] (blkA) -- ++ (-0.55,0) |- (blkB);
  \draw[very thick,->] (init) -- (1,-3.80);
  \node[anchor=south west,omegas] at (1,-3.8) {$\Omega_2$};
  \node[anchor=north west,omegas] at (1,-3.8) {$\Omega_2$};
  \node[anchor=north east,omegas] at (blkB.south) {$\Omega_2$};
  \node[anchor=north west,omegas] at (flush.south) {$\Omega_2$};
  \node (replstart) at (1, -3.80) {};
  \draw[->] (1,-3.80) -- (blkA);
  \draw node[align=right,anchor=west] at (0.45,-5.05) {bypass};
  \node[anchor=west,omegas] at (0.45,-4.75) {$\Omega_3$};
  \draw node[align=left,anchor=west] at (1.4,-5.15) {back-edge};
  \node[anchor=west,omegas] at (1.4,-4.85) {$\Omega_3$};
  \draw node[align=right,anchor=east] at (2.25,-4.35) {\lstinline!replicate!\\(line 18)};
  \draw[dashed,->] (storeB) -- (blkB);
  \draw[->] (blkB) -- (1,-6.10);
  \node (replend) at (1, -6.10) {};
  \draw[very thick,->] (1,-6.10) -- (flush);
  \draw[very thick,->] (fetchA) -- (storeA) node[midway,above,omegas] {$\Omega_2$};
  \draw[very thick,->] (fetchC) -- (storeC) node[midway,below,omegas] {$\Omega_2$};
  \draw node[align=left,anchor=north west] at (-0.45,-2.5) {\lstinline!foreach!\\(line 12)};
  \draw node[align=left,anchor=north west] at (-0.60,-1.3) {\lstinline!foreach!\\(line 6)};
  \draw[very thick,->] (1,-2.5) -- (init);
  \draw[very thick,->] (flush) -- (1,-7.6);
  \node[anchor=north west,omegas] at (1,-7.6) {$\Omega_1$};
  \draw[->] (1,-7.6) |- (1,-7.75) -| (fetchC);

  \draw[dashed,->] (storeA) |- (1,-2.25) -| (1,-2.5);
  \draw[->] (in) |- (1,-2.25) -| (1,-2.5);
  \draw[->] (in) |- (fetchA);
  \draw[->] (in) |- (-0.525,-2.25) |- (0,-7.75) -| (fetchC);
  \draw[dashed,->] (storeC) -- (1.0, -8.8);
  \draw[dashed,->] (1.0, -8.8) -- ++ (0, -0.4);
  \node[anchor=north west,omegas] at (1,-8.8) {$\Omega_0$};

  \node[anchor=east] (void) at (2.6, -7.85) {void};
  \node[anchor=east] (scalar) at (2.6, -8.2) {scalar};
  \node[anchor=east] (vector) at (2.6, -8.55) {vector};
  \draw[dashed,->] (void) ++ (-0.6,0) -- (void);
  \draw[->] (scalar) ++ (-0.6,0) -- (scalar);
  \draw[very thick,->] (vector) ++ (-0.6,0) -- (vector);

\end{tikzpicture}
  \caption{Basic dataflow for \Cref{fig:motivate} showing hierarchical parallel sections and access-pattern-optimized memories, as specified by the input program before optimization. In \name{} code, control-flow \emph{becomes} dataflow, so dataflow arrows represent control operations.}\label{fig:motiv_dataflow}
\end{figure}
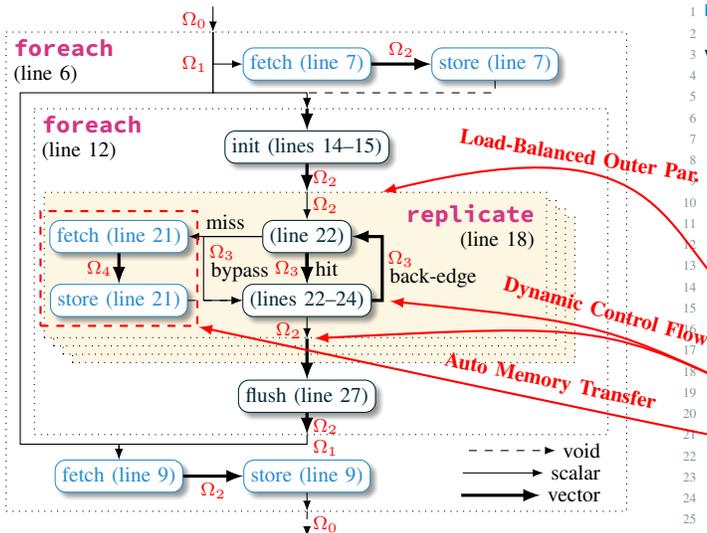

Finally, \name{} supports dynamic thread spawning and termination using the \lstinline!fork! construct, unlike GPUs that spawn threads in rigid blocks at kernel launch time.
While \lstinline{foreach} creates threads beneath a parent, \lstinline{fork} creates new threads at the same hierarchy level.
This is similar to a POSIX fork~\cite{posixfork}, except \name{} can fork an arbitrary number of threads. 
We also provide an \lstinline!exit! operation that terminates thread execution without returning a value to the reduction.

\lstinline!fork! is generally more expensive than \lstinline!foreach!: the latter can use scalar-to-vector broadcasting to reduce on-chip dataflow requirements, while the former must duplicate all live variables because no hierarchy is added.
However, \lstinline!foreach! loops have static restrictions on branching---a child thread cannot branch outside the \lstinline!foreach!.
Because \lstinline!fork! does not add hierarchy, there are no such restrictions.

\paragraph{Access-Pattern Optimized Memories}
Memory access patterns that are easy for programmers are hard for hardware and vice versa, and this dichotomy is most evident when programmers are writing sequential code.
Programmers would rather access memory one word at time, relying on hardware like caches to keep access times low.
Conversely, hardware prefers loading an entire vector from DRAM into an SRAM scratchpad and then accessing the scratchpad explicitly.
We balance the burden of achieving high performance in a scratchpad-based design between the programmer and the compiler using the semantics shown in \Cref{tab:adapters}.

The programmer starts by choosing an appropriate access mode. 
For example, affine accesses with known dimensions should use views, which coordinate large-tile transfers and can be accessed within \lstinline{foreach} loops.
Conversely, data-dependent sequential accesses should use iterators, which our compiler maps to optimized hardware that coordinates small-block transfers dynamically for each thread based on local control-flow decisions, as shown in \Cref{fig:hitmiss}.

To support these primitives, and maintain Spatial's~\cite{koeplinger2018spatial} support for multi-buffered on-chip SRAM, \name{} supports dynamic allocation of on-chip memories.
Like Spatial, \name{} requires that on-chip SRAMs have a compile-time fixed size.
However, unlike Spatial and SARA~\cite{zhang2021sara}, \name{} supports out-of-order allocation and deallocation to enable reordered thread execution.

\begin{figure}
  \vskip-1em
  \begin{tikzpicture}[overlay,x=1cm,y=1cm]%
  \filldraw [fill=solarized-base3,draw=solarized-base2] (1.30,-4.80) rectangle (8.35,-6.85);%
\end{tikzpicture}%
\begin{lstlisting}[style=revetsmall,escapeinside={(*@}{@*)}]
DRAM<char> input;  DRAM<int> offsets;  DRAM<int> lengths; 

void main(int count) {
  foreach (count by 1024) { int outer =>
    (*@\tikzmark{autoread}@*)ReadView<1024> in_view(offsets, outer);
    // in_view is loaded here
    (*@\tikzmark{autoflush}@*)WriteView<1024> out_view(lengths, outer);
    // This foreach will be rewritten
    // as a hierarchy-less fork statement.
    foreach (1024) { int idx =>
      pragma(eliminate_hierarchy);
      int len = 0;
      int off = in_view[idx];
      // Transition from vector dataflow to scalar
      // dataflow that is internally vectorized.
      (*@\tikzmark{outerpar}@*)replicate (4) {
        // Stateful control within parallel regions, 
        // including stateful updates to the iterator.
        ReadIt<64> it(input, off);
        (*@\tikzmark{controldecision}@*)while (*it) { 
          len++;
          (*@\tikzmark{dynreload}@*)it++; // Iterator is dynamically reloaded.
        };
      };
      out_view[idx] = len;
    };
    // out_view is flushed here.
  };
}
\end{lstlisting}
  \tikzstyle{arrow}=[<->,thick,draw=red]
  \tikzstyle{arrownode}=[above,midway,sloped,font={\bfseries\footnotesize\color{red}},rounded corners=1ex]
  \begin{tikzpicture}[overlay]%
    \node[thick,draw=red, dashed, fit=(fetchB) (storeB)] (tmp) {};
    \draw[arrow] (dynreload) .. controls ++ (-2.5, 0.5) .. node[arrownode,pos=0.6] {Auto Memory Transfer} (tmp.south east);%
    \draw[arrow] (controldecision) .. controls ++ (-2.5, 1.2) .. node[arrownode,pos=0.50] {Dynamic Control Flow} (be-turn);%
    \draw[arrow] (controldecision) .. controls ++ (-2.5, 1.2) .. (replend);%
    \draw[arrow] ([yshift=0.7ex]outerpar)  .. controls ++ (-1.5,1.0) and ++ (4,1) .. node[arrownode,pos=0.70,anchor=south] {Load-Balanced Outer Par.} ($(replstart)+(1cm,0)$);%
\end{tikzpicture}
\vskip-1.2\baselineskip
\lstset{style=revetfootnote}
\caption{\name{} code for \lstinline!strlen()!. Code in the highlighted box could not be expressed in Spatial~\cite{koeplinger2018spatial}. The simplicity of \name's programming model is highlighted by red arrows, which show how standard imperative language features map to new dataflow features. }
\label{fig:motivate}
\end{figure}

\subsection{Case Study: \lstinline{strlen}}
\Cref{fig:motiv_dataflow} shows how \name's parallel constructs work together to map the \lstinline!strlen! computation in \Cref{fig:motivate} across spatially distributed, parallel pipelines.
Specifically, the body of the \lstinline!while! loop on line 20  and the implicit fill path for the \lstinline!ReadIt! are critical code sections, so \name{} uses explicit parallelism to run them on multiple vector pipeline.
The outer \lstinline{foreach} (line 10) will first transform a scalar value into a vector of threads. 
The \lstinline{replicate} (line 16) will use outer-loop (non-vector) parallelism to distribute those threads across the chip as multiple scalar pipelines (instead of one vector pipeline).
Next, the \lstinline{while} statement automatically adds vector parallelism again within each inner pipeline, because threads are executing independently on their own lanes.
Finally, the \lstinline!ReadIt! transfer path is scalar (refills are infrequent), so the implicit \lstinline!foreach! inside it can be vectorized again.
\name's flexible programming  model lets the compiled dataflow code transition between vector and scalar execution without explicit programmer intervention.

\begin{table}
  {
    \centering
      \lstset{style=revetfootnote}
    \caption{On-chip memory adapters. Only array-decay memories can be allocated outside a \lstinline!foreach! loop and accessed inside.}
    \footnotesize
    \lstset{style=revetsmall}
    \label{tab:adapters}    
    \begin{tabular}{p{1.7in}ccc}
    \toprule
        Access Pattern \& Name & Read & Write & Array-Decay \\
    \midrule
        \quad \lstinline!SRAM<type,size>()!  & Yes & Yes & Yes\\
      \multicolumn{4}{p{2.6in}}{\vskip-0.4\baselineskip Small auto-fetched and -stored tiles:} \\
        \quad \lstinline!ReadView<size>(dram, base)!  & Yes & & Yes\\
        \quad \lstinline!WriteView<size>(dram, base)!  &  & Yes&Yes\\
        \quad \lstinline!ModifyView<size>(dram, base)!   & Yes & Yes&Yes\\
        \multicolumn{4}{p{2.6in}}{\vskip-0.4\baselineskip Linear read, optionally with peek \lstinline!tile! elements ahead:}\\
        \quad\lstinline!ReadIt<tile>(dram, seek)! & Yes & & \\
        \quad\lstinline!PeekReadIt<tile>(dram, seek)!  & Yes & & \\
        \vskip-0.5\baselineskip Linear write (output iter): \\
        \quad\lstinline!WriteIt<tile>(dram, seek)!  && Yes & \\
        \multicolumn{4}{p{2.6in}}{\vskip-0.4\baselineskip Linear write with manual flush. May overwrite a word where a sub-word is touched:}\\
        \quad\lstinline!ManualWriteIt<tile>(dram, seek)!  &  &Yes & \\
    \bottomrule
    \end{tabular}
  }
\end{table}

\section{Implementation}
\label{sec:implementation}
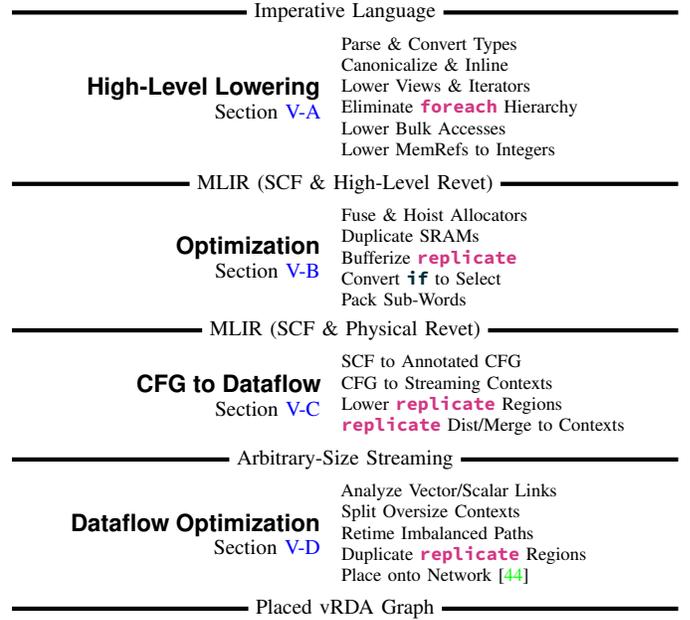
\begin{figure}
  \centering
  \footnotesize
  \lstset{style=revetsmall}

  \begin{tikzpicture}[auto,node distance=0.0\baselineskip,start chain=going below,box/.style={inner sep=4pt},rbox/.style={box,align=left,anchor=east},lbox/.style={box,align=right,anchor=west},y=\baselineskip]
    \newcommand{\IRDef}[1]{{
      \node[on chain] (tmpXX) {#1};
      \draw[ultra thick] (tmpXX) ++ (-0.5\columnwidth,0) -- (tmpXX);
      \draw[ultra thick] (tmpXX) -- ++ (0.5\columnwidth,0);
    }}
    \newcommand{\PassDef}[3]{{
      \node[on chain,align=center] (tmpXX) {\parbox{1.35in}{\raggedleft {\bfseries\sffamily\small #1}\\#2}\hskip1em\parbox{1.5in}{\scriptsize #3}};
    }}
    \IRDef{Imperative Language};
    \PassDef{High-Level Lowering}{\Cref{sec:frontendlowering}}{Parse \& Convert Types\\Canonicalize \& Inline\\Lower Views \& Iterators \\Eliminate \lstinline!foreach! Hierarchy\\Lower Bulk Accesses\\Lower MemRefs to Integers};
    \IRDef{MLIR (SCF \& High-Level \name{})};
    \PassDef{Optimization}{\Cref{sec:optimization}}{Fuse \& Hoist Allocators\\Duplicate SRAMs\\Bufferize \lstinline!replicate!\\Convert \lstinline!if! to Select\\Pack Sub-Words};
    \IRDef{MLIR (SCF \& Physical \name{})};
    \PassDef{CFG to Dataflow}{\Cref{sec:cflowdflowimpl}}{SCF to Annotated CFG\\CFG to Streaming Contexts\\Lower \lstinline!replicate! Regions\\\lstinline!replicate! Dist/Merge to Contexts};
    \IRDef{Arbitrary-Size Streaming};
    \PassDef{Dataflow Optimization}{\Cref{sec:streamingtransforms}}{Analyze Vector/Scalar Links\\Split Oversize Contexts\\Retime Imbalanced Paths\\Duplicate \lstinline!replicate! Regions\\Place onto Network~\cite{zhang2019scalable}};
    \IRDef{Placed vRDA Graph};
\end{tikzpicture}
\vskip-.5\baselineskip
  \caption{\name's compiler passes and IRs.}
\label{fig:compilerstages}
\end{figure}

In this section, we describe the practical details behind our prototype implementation, which follows the stages shown in \Cref{fig:compilerstages}. 
\name{}'s compiler starts by parsing the language and eliminating several constructs implemented to improve programmer productivity, including views and iterators.
Then, the compiler performs optimizations to improve the efficiency of the generated dataflow.
At this point, the IR still contains constructs like \lstinline!while! loops and \lstinline!if! statements, but it has been rewritten to be physically realizable (e.g., bulk memory accesses have been converted to \lstinline!foreach! loops) and to generate a more optimal dataflow output (e.g., small \lstinline!if! statements have been converted to predication).
Then, the compiler lowers the CFG representation to a dataflow format---effectively, this consists of rewriting basic blocks as infinitely large virtual CUs and replacing structured control flow constructs with the corresponding \name{} primitives discussed in \Cref{sec:tensorprimitives}.
Finally, the arbitrary-size virtual CUs are split so that they meet the physical constraints (operation count, input/output count) of our vRDA backend.

\subsection{Front-End Lowering}
\label{sec:frontendlowering}
\name{} uses a front-end~\cite{parr2013definitive} that emits code into an MLIR~\cite{lattner2020mlir} representation as a mixture of the structured control flow dialect (SCF) and a custom \name{} dialect that captures our custom front-end features (\Cref{sec:lang}).
Our front-end also inserts type conversion operations as needed.
Then, we progressively lower the high-level \name{} memory operations (e.g., iterators) until every memory is expressed as an SRAM with scalar accesses and perform hierarchy elimination if requested.

\paragraph{View \& Iterator Lowering}
We rewrite \name{} views and iterators (\Cref{tab:adapters}) into MemRefs (MLIR's annotated memory type) and integers.
Views are simple: allocations are replaced with a MemRef allocation and a bulk load (if needed) and deallocations are replaced with a MemRef deallocation and a bulk store (if needed).
These primitives are more efficient for sub-word types because a backend-inserted bulk store can process 32 bits per cycle.

Iterators are slightly more complicated: the basic \lstinline!ReadIt!, for example, has a MemRef buffer, a local pointer (8 bits to reduce dataflow overhead), and a global pointer (in SRAM).
The global pointer is fetched and incremented only when the local pointer wraps around.
Because dereference is less common, we fill read iterators' buffers only at dereference to decrease the amount of hardware mapped.
\lstinline!WriteIt!s can be flushed at increments or deallocation, which would na\"ively require two store paths. 
To avoid this, the \lstinline!ManualWriteIt! takes an input at increment indicating the last loop iteration to elide the deallocation flush.

\paragraph{Foreach Hierarchy Elimination}
\lstinline!foreach! regions are eventually lowered to streaming tensor operations (\Cref{sec:tensorprimitives}), which use barriers to sequence threads.
However, barriers can limit parallelism inside \lstinline!while! loops, where they force a total flush of the loop body before new threads can enter the loop.
Because barriers change side-effect ordering, they cannot be automatically eliminated, so we only rewrite \lstinline{pragma}-annotated \lstinline!foreach! statements.

\begin{figure}
  \centering
\begin{minipage}[t]{0.4\columnwidth}
\begin{lstlisting}[style=revetsmall]
foreach (n) { 
  // loop body
  int i => ... 
}
\end{lstlisting}
\centering
\begin{tikzpicture}[auto, node distance=9mm, y=0.70cm,
  box/.style = {draw,rounded corners,blur shadow,fill=white, align=center}]
  \footnotesize
  \draw[fill=white,dotted] (-2,1) -| (0.5,-1) -| cycle;
  \draw node[align=left,anchor=north west] at (-2,1) {\lstinline!foreach!};
  \node[box] (body) at (0,0) {body};
  \draw[very thick,->] (0,1) -- (body);
  \draw[very thick,->] (body) -- (0,-1);
\end{tikzpicture}
\end{minipage}%
\begin{minipage}[t]{0.6\columnwidth}
\begin{lstlisting}[style=revetsmall]
mem[ptr] = n;
int i = fork(n);
... // loop body
if (mem[ptr]--) {
  exit();
};
\end{lstlisting}
\centering
\begin{tikzpicture}[auto, node distance=9mm, x=1.2cm, y=0.6cm,
  box/.style = {draw,rounded corners,blur shadow,fill=white, align=center}]
  \footnotesize
  \node[box] (body) at (0,1) {body};
  \node[box] (fork) at (-0.9,1) {\lstinline[style=revetsmall]!fork!};
  \node[box] (st) at (-1.7,1) {store};
  \node[box] (mem) at (0,0) {memory};
  \node[box] (dec) at (1.1,1) {dec/\lstinline[style=revetsmall]!exit!};

  \draw[->] (st) ++ (-0.5,0) -- (st);
  \draw[->] (st) -- (fork);
  \draw[very thick,->] (fork) -- (body);
  \draw[very thick,->] (body) -- (dec);
  \draw[->] (st) |- (mem);
  \draw[very thick,<->] (dec) |- (mem);
  \draw[very thick,<->] (dec) |- (mem);
  \draw[->] (dec) -- ++ (0.65,0);
\end{tikzpicture}
\end{minipage}
\vskip.6\baselineskip
\footnotesize
\begin{minipage}[t]{0.4\columnwidth}
  \centering
(a) Hierarchy
\end{minipage}%
\begin{minipage}[t]{0.6\columnwidth}
  \centering
(b) Flattened
\end{minipage}
\vskip-0.2\baselineskip
\lstset{style=revetfootnote}
  \caption{Hierarchy elimination (\lstinline!foreach! to \lstinline!fork!).}\label{fig:flatten}
\end{figure}
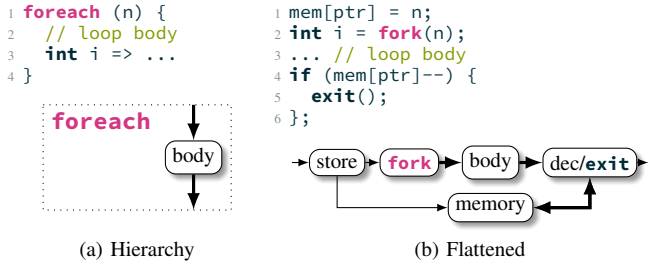

When rewriting these statements, we initialize a memory location with the number of elements expected and execute a \lstinline!fork! to create hierarchy-less threads, as shown in \Cref{fig:flatten}.
Instead of reduction, threads atomically fetch and decrement the shared memory location.
If zero elements remain, the thread is the last one and iteration is complete; otherwise, the thread exits.
This removes the strict synchronization between \lstinline!foreach! loops that would otherwise be imposed by SLTF bariers: when using \lstinline!fork! statements, the straggling children of one parent can be interleaved with those of the next parent.

\subsection{Optimization}
\label{sec:optimization}
At this point, our IR is in a mix of SCF, standard arithmetic operations, and physical memory operations.
We run several rewrite passes before lowering the code to dataflow, in addition to existing MLIR passes.
These passes rewrite the high-level IR to increase dataflow performance.

\paragraph{SRAM Allocator Optimizations}
\label{sec:fusealloc}
To avoid fragmentation, \name{}'s on-chip allocation relies on compile-time-determined fixed-size buffers at each memory.
For example, if an SRAM buffer is specified as \SI{64}{B} (matching the vector width), we rewrite every memory access as: $\mathrm{ptr}\times\mathrm{64}+\mathrm{off}.$ 
This transformation means that every integer within a range $[0,\mathrm{max})$ is a valid pointer, and one pointer can be used at multiple memories as long as it is in range.
By default, the maximum pointer is the size of a single MU divided by the thread-local buffer size, but users can increase the thread count using a \lstinline!pragma!, which will cause multiple MUs to be inserted to increase storage.

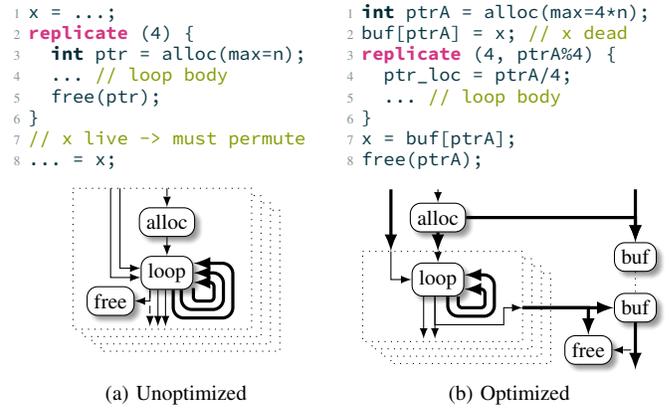
\begin{figure}
  \centering
\begin{minipage}[t]{0.5\columnwidth}
\begin{lstlisting}[style=revetsmall]
x = ...;
replicate (4) { 
  int ptr = alloc(max=n);
  ... // loop body
  free(ptr);
}
// x live -> must permute
... = x;
\end{lstlisting}
\centering
\begin{tikzpicture}[auto, node distance=9mm, x=2.5cm, y=0.75cm,
  box/.style = {draw,rounded corners,blur shadow,fill=white, align=center}]
  \footnotesize
  \foreach\x in {3,2,1,0}
    \draw[fill=white,dotted] (-0.50+\x*0.05,0.5-\x*0.13) -| (0.45+\x*0.05,-2-\x*0.13) -| cycle;
  \node[box] (alloc) at (0,-0.1) {alloc};
  \node[box] (free) at (-0.3,-1.5) {free};
  \node[box] (loop) at (0,-1) {loop};
  \draw[dashed,->] (0,0.5) -- (alloc);
  \draw[->] (alloc) -- (loop);
  \draw[->] (-0.25,0.5) |- (loop.170);
  \draw[->] (-0.30,0.5) |- (loop.190);
  \draw[dashed,->] (loop.225) -- ++(0,-0.65);
  \draw[->] (loop.225) |- (free);
  \draw[->] (loop.242) -- ++(0,-0.65);
  \draw[->] (loop.265) -- ++(0,-0.65);
  \draw[rounded corners=1.5pt,very thick,->] (loop.330) |- (0.25,-1.5) |- (loop.340);
  \draw[rounded corners,very thick,->] (loop.310) |- (0.30,-1.65) |- (loop.360);
  \draw[rounded corners,very thick,->] (loop.290) |- (0.35,-1.8) |- (loop.20);
\end{tikzpicture}
  \footnotesize
\end{minipage}%
\begin{minipage}[t]{0.5\columnwidth}
\begin{lstlisting}[style=revetsmall]
int ptrA = alloc(max=4*n);
buf[ptrA] = x; // x dead
replicate (4, ptrA%4) {
  ptr_loc = ptrA/4;
  ... // loop body
}
x = buf[ptrA];
free(ptrA);
\end{lstlisting}
\centering
\begin{tikzpicture}[auto, node distance=9mm, x=2.5cm, y=0.75cm,
  box/.style = {draw,rounded corners,blur shadow,fill=white, align=center}]
  \footnotesize
  \foreach\x in {3,2,1,0}
    \draw[fill=white,dotted] (-0.40+\x*0.05,-0.5-\x*0.13) -| (0.45+\x*0.05,-2.1-\x*0.13) -| cycle;
  \node[box] (alloc) at (0,0.1) {alloc};
  \node[box] (loop) at (0,-1) {loop};
  \node[box] (bufA) at (1.05,-0.6) {buf};
  \node[box] (bufB) at (1.05,-1.5) {buf};
  \node[box] (free) at (0.80,-2.25) {free};
  \draw[dashed,->] (0,0.6) -- (alloc);
  \draw[very thick,->] (alloc) -- (0,-0.5);
  \draw[very thick,->] (-0.25,0.6) -- (-0.25,-0.5);
  \draw[very thick,->] (1.05,0.6) -- (bufA);
  \draw[very thick,->] (alloc) -| (bufA);
  \draw[very thick,->] (bufB) -- ++(0,-1.1);
  \draw[very thick,->] (0.45,-1.5) -- (bufB);
  \draw[very thick,->] (0.45,-1.5) -| (free);
  \draw[dashed,->] (bufB) |- (free);
  \draw[->] (0,-0.5) -- (loop);
  \draw[->] (-0.25,-0.5) |- (loop.180);
  \draw[->] (loop.230) -- ++(0,-0.75);
  \draw[->] (loop.260) -- ++(0,-0.75);
  \draw[->] (loop.260) -- ++(0,-0.50) -| (0.35,-1.5) -- (0.45,-1.5);
  \draw[dotted] (bufA) -- (bufB); 
  \draw[rounded corners=1.5pt,very thick,->] (loop.320) |- (0.25,-1.5) |- (loop.340);
  \draw[rounded corners,very thick,->] (loop.300) |- (0.30,-1.65) |- (loop.370);
\end{tikzpicture}
\footnotesize
\end{minipage}
\footnotesize
\vskip.6\baselineskip
\begin{minipage}[t]{0.5\columnwidth}
  \centering
(a) Unoptimized
\end{minipage}%
\begin{minipage}[t]{0.5\columnwidth}
  \centering
(b) Optimized
\end{minipage}
\vskip-0.2\baselineskip
\lstset{style=revetfootnote}
  \caption{Allocator hoisting outside and buffering of a live value around a \lstinline!replicate!. The low bits of the hoisted pointer steer threads to a \lstinline!replicate! region, and the high bits are used within it.}\label{fig:hoistbuffer}
\end{figure}

Allocation fusion lowers the number of pointers that must be tracked in dataflow.
We fuse all allocations in a basic block, taking the intersection of valid pointers for the memories to be fused.
Because each allocator is sampling from a range defined by a single maximum value, the fused range is defined by the smallest maximum pointer across all memories in the basic block.
Finally, \name{} loads these pointers into a queue stored in a memory unit, so allocation pops a pointer from this queue and deallocation pushes it back.
\label{sec:hoistalloc}
\paragraph{Allocator Hoisting \& Bufferization}
\label{sec:bufferize}
If a \lstinline!replicate! region contains one allocation after fusion, we can increase its range, using the low bits to point to a specific region and the high bits to address an SRAM buffer within that region.
This has two benefits. 
First, it lowers resource requirements by vectorizing allocation (\Cref{fig:hoistbuffer}), with one allocator globally instead of one per region.
Second, it provides native round-robin load balancing: regions only receive new allocations after they complete existing ones.

\lstinline!replicate! regions take advantage of \name's unordered-threads abstraction, and do not maintain ordering.
Therefore, when a thread enters a \lstinline!replicate!, all of its live values would have to be sent {into} the \lstinline!replicate!, even if they are not used inside it, creating excessive network congestion.
Instead, we \emph{reuse} the single live pointer into a \lstinline!replicate! (if one has been hoisted) to bufferize live values around it, inserting an SRAM to store the value (\Cref{fig:hoistbuffer}).
Then, we replace all uses after the replicate with a load from this SRAM, so the value is not live through the replicate.

\paragraph{If-to-Select Conversion}
\label{sec:lowerif}
Na\"ive dataflow would assign a compute unit to each branch of an \lstinline{if} statement, but \lstinline{if} statements without inner loops would just leave empty lanes.
Therefore, we inline all \lstinline{if} statements that lack inner loops, replacing them with conditional moves and predicating memory operations.
This is more powerful than MLIR's default of only rewriting empty \lstinline{if}s.

\paragraph{Sub-Word Packing}
\label{sec:subwordpack}
Every variable that is live into a merge operation consumes a significant number of network resources and input buffers, leading to congestion.
Therefore, we identify sub-word values (\lstinline{int8} and \lstinline{int16}) that are live into or out of \lstinline{while} loops.
In a na\"ive dataflow lowering, each of these would have to be promoted to an \lstinline{int32} because threads execute on 32-bit lanes. 
However, this would waste buffers and on-chip links, which are critical resources when mapping.
Therefore, we pack these values into a single \lstinline{int32}, making sure to minimize permutation for nested loops.
We also optimize AND, shift, and OR operations that can be expressed as sub-word reads and writes on fixed boundaries.

\subsection{CFG to Dataflow Lowering}
\label{sec:cflowdflowimpl}
In this section, we describe a prototype system that operates on MLIR's SCF dialect using specially-annotated CFGs as an intermediate format to ensure our hierarchical CFG's constraints are respected.

\paragraph{Graph Annotation}
Between SCF and dataflow, we use an annotated CFG that indicates which edges should be forward merges and which edges should be forward-backward merges using specialized block terminators.
We also flatten \lstinline!foreach! regions into the main CFG and replace their input and output edges with special counter and reduce terminators (like Tapir~\cite{schardl2017tapir}).
We ensure that every block has no more than two predecessors to meet our hardware's merge constraint.

\paragraph{Basic-Block Inputs}
When mapping a block, we start by identifying all live-in variables and determining whether they should be broadcast using a mapping table of blocks to nesting depths.
For example, a block following an \lstinline!if! statement will have two sets of live-ins (one from each branch, followed by a merge operation) in addition to broadcasts.

In case there are no live-in values, a data-less void value is inserted between blocks to guarantee a correct number of live-outs.
This void value is chained through pipelined memory operations to guarantee a data dependence to enforce ordering after contexts have been split.
After cloning inputs, we clone the basic block as element-wise operations.

\paragraph{Basic-Block Outputs}
Basic-block live-out values are mapped based on the terminator operation.
Unconditional branches map to unconditional outputs: each lane in the pipeline is sent as an output.
Conditional branches map to filter outputs, taking a condition and a value per lane; only lanes matching the condition are sent.
\lstinline!foreach! block terminators map to counters, which are later moved to the head of their destination context, and reductions map to a reduction operation at the end of the context pipeline.
Finally, outputs exiting a while-loop region strip hierarchy without reduction.

\paragraph{Replicate}
The logic to connect \lstinline!replicate! regions to the enclosing CFG is implemented using the filter and merge primitives shown in \Cref{sec:dataflow}: each thread is broadcast to a filter before every contained CFG and only threads with matching keys are forwarded.
Finally, return values from the contained CFG are merged using a tree of forward merge units.

To speed compilation, we use late unrolling for \lstinline!replicate! regions: we create one node with code and multiple nodes for references, which are duplicated immediately before placement.
We then insert work-distribution and output-merging logic using the filter and forward-merge primitives from \Cref{sec:dataflow}.
To avoid a single slow \lstinline!replicate! region stalling a hoisted allocator and starving faster regions, we insert link-retiming buffers in the work-distribution logic.

\subsection{Dataflow Optimization}
\label{sec:streamingtransforms}
In the previous subsection, we described lowering to a \emph{virtual} streaming IR, which is a one-to-one mapping of control-flow constructs to dataflow.
Here, we describe passes that transform the virtual streaming IR into a \emph{physical} streaming IR. 
These include passes like splitting that ensure our IR can map to on-chip units and optimization passes like retiming.

\paragraph{Vector/Scalar Link Analysis}
Because some buffers can only store scalars, accurately mapping virtual links to either vector or scalar physical links is important---especially for merges.
Only two vector-vector merges fit in a context (a total of four vector buffers), but four scalar-vector merges fit, halving the number of resources required.
However, if a high-throughput link is mapped to a scalar physical link, then its throughput will be only $\rfrac{1}{16}$ of peak.
Therefore, we treat links as vector by default, except blocks following \lstinline!while! loops and the entrances and exits of \lstinline!replicate! regions and the main program graph.
However, a \lstinline{pragma} can override this.

\paragraph{Splitting, Retiming, \& Placement}

Initially, compute operations are mixed with memory operations in contexts, and a single context may have memory operations at two or more memories and an impossible number of inputs, outputs, or operations.
We first place every memory operation into its own context, and then split over-size contexts.
We next insert buffers to avoid deadlock 
\ifdefined\longversion
(\Cref{sec:deadlock}) 
\fi
and mitigate path-delay imbalances~\cite{zhang2021sara}.
Finally, we place the partitioned graph using previously proposed tools~\cite{zhang2019scalable}, prioritizing deeply nested nodes.

\begin{table}
  \footnotesize
  \centering
  \caption{RDA parameters used in our evaluation.}\label{tab:params}
  \begin{tabu}{Xr}
    \toprule
     Compute units (200) & 16 lanes, 6 stages, 6 vec/scal regs/lane/stage\\
     Memory units (200)  & 16 banks, \SI{256}{KiB} total\\
    Buffers  (per unit)       & 4\texttimes256 word vec.,  4\texttimes64 word scal.  \\
    Outputs   (per unit)      & 4 vector, 4 scalar\\
    Network        & 3\texttimes{} vector, 6\texttimes{} scalar, dynamic            \\
    DRAM        & HBM2,    $\sim$\SI{900}{GB/s}, 32B burst            \\
      \bottomrule

  \end{tabu}
\end{table}

\begin{table*}
  \centering
  {
  \centering
  \lstset{style=revetfootnote}
  \caption{Applications and data distributions used to test \name{}. Key features are shown; applications use additional features (e.g., \lstinline!replicate! and \lstinline!WriteView! for data storage).  $^\dagger$This application ran slightly \emph{slower} for larger datasets (1.4\texttimes{}).}\label{tab:apps_revet}
  \lstset{style=revetsmall}
  \footnotesize
  \vskip-.3\baselineskip
  \begin{tabu}{X@{\hskip1ex}rlll@{\hskip1ex}rrr}
    \toprule
                &       &                       &                                                     &                                                         & \multicolumn{3}{c}{Scale (MiB)} \\ \cmidrule{6-8}
                & Lines & Description           & Per-Thread Dataset                                  & Key Features                                            & \name                            & GPU                                  & CPU   \\ \midrule
     isipv4     & 34    & DFA regex             & 90\% valid addresses, 10\% `INVALID'                & \lstinline!replicate! (\texttimes2)                     & 38                               & 1527                                 & 15275 \\
     ip2int     & 41    & Parsing               & Random IPv4 addresses                               & \lstinline!replicate! (\texttimes2)                     & 52                               & 2070                                 & 20700 \\
     murmur3    & 62    & Data hashing          & \SI{64}{B} blobs                                    & \lstinline!ReadIt!                                      & 136                              & 1360                                 & 13600 \\
     hash-table & 56    & Hash-table lookup     & int32 keys/values, $10^8$ slots, 25\% load          & \lstinline!ReadIt!                                      & 16                               & 2400                                 & 24000 \\
     search     & 54    & Exact-match search    & Find `Moby Dick', \SI{256}{B} chunks of `Moby Dick' & \lstinline!PeekReadIt!, \lstinline!while! (\texttimes2) & 260                              & 41\makebox[0pt]{\hskip1ex$^\dagger$} & 26000 \\
     huff-dec   & 40    & Decompression         & 64 codes, 16-bit max length                         & \lstinline!ReadIt!                                      & 171                              & 1714                                 & 17140 \\
     huff-enc   & 58    & Compression           & 64 codes, 16-bit max length                         & \lstinline!ManualWriteIt!                               & 171                              & 1714                                 & 17140 \\
     kD-tree    & 74    & Count points in rect. & $10^8$-point grid, random searches yield 16 points  & \lstinline!fork!                                        & 40                               & 800                                  & 8000  \\
     \bottomrule
  \end{tabu}} 
\end{table*}

\begin{table*}
  \scriptsize
  \vskip-.4\baselineskip
  \setlength{\tabcolsep}{5.4pt}
  \centering
  \caption{Resources used by \name{} applications.}\label{tab:resources}
  \vskip-.3\baselineskip
  \begin{tabu}{Xrrrrrrrrrrrrrrrrrrr}
    \toprule
    & \multicolumn{2}{c}{Parallelization} & \multicolumn{3}{c}{Inner} & \multicolumn{3}{c}{Outer} & \multicolumn{2}{c}{Replicate} & \multicolumn{3}{c}{Retime/Buffer (MU)} & \multicolumn{3}{c}{Total} & \multicolumn{3}{c}{HBM2 (\%)}\\ 
    \cmidrule(lr){2-3}
    \cmidrule(lr){4-6}
    \cmidrule(lr){7-9}
    \cmidrule(lr){10-11}
    \cmidrule(lr){12-14}
    \cmidrule(lr){15-17}
    \cmidrule(lr){18-20}
               & Outer         & Lanes & CU  & MU & AG & CU & MU & AG & CU & MU & Deadlock & Buffer & Retime & CU  & MU  & AG & Read & Write & Total \\
    \midrule                                                                                          
    isipv4     & 3\texttimes9  & 432   & 81  & 54 & 27 & 23 & 10 & 6  & 43 & 10 & 0        & 12     & 73     & 147 & 159 & 33 &  83.0&	0.5&	83.5\\
    ip2int     & 3\texttimes10 & 480   & 90  & 30 & 30 & 23 & 10 & 6  & 46 & 10 & 0        & 12     & 79     & 159 & 141 & 36 &  68.5&	13.1&	81.6\\
    murmur3    & 14            & 224   & 112 & 28 & 14 & 11 & 5  & 3  & 21 & 7  & 42       & 5      & 20     & 144 & 107 & 17 &  73.9&	4.1&	78.0\\
    hash-table & 16            & 256   & 112 & 32 & 16 & 13 & 4  & 2  & 23 & 6  & 48       & 4      & 22     & 148 & 116 & 18 &  29.6&	2.3&	32.0\\
    search     & 8             & 128   & 120 & 40 & 8  & 10 & 4  & 2  & 12 & 3  & 32       & 4      & 13     & 142 & 96  & 10 &  66.3&	0.8&	67.1\\
    huff-dec   & 9             & 144   & 135 & 45 & 18 & 7  & 3  & 1  & 13 & 4  & 27       & 2      & 41     & 155 & 122 & 19 &  17.1&	31.6&	48.7\\
    huff-enc   & 9             & 144   & 126 & 45 & 18 & 10 & 4  & 2  & 13 & 4  & 27       & 4      & 43     & 149 & 127 & 20 &  35.0&	17.5&	52.5\\
    kD-tree    & 5             & 80    & 110 & 55 & 65 & 3  & 1  & 0  & 7  & 2  & 10       & 0      & 36     & 120 & 104 & 65 &  57.1&	0.2&	57.3\\
    \bottomrule
  \end{tabu}
\end{table*}

\section{Evaluation}\label{sec:eval}
After discussing our methodology, we discuss \name's performance and how optimizations improve generated code and out-perform industrial baselines on a variety of applications.

\subsection{Methodology}
We evaluate \name{} using a cycle-accurate vRDA simulation including a model of HBM2 memory~\cite{kim2015ramulator,standard2013high,zhang2019scalable} against real-world baseline designs on a variety of kernels.

\paragraph{Hardware Model}
To evaluate the dataflow threads programming model and backend abstraction in a physically-constrained environment, we use an abstract machine model based on Plasticine~\cite{prabhakar2017plasticine}.
Our abstract vRDA comprises 200 compute units (CUs), 200 memory units (MUs), and 80 DRAM address generators (AGs) connected by a flexible on-chip network~\cite{zhang2019scalable}; the parameters we use are shown in \Cref{tab:params}.
Our vRDA backend uses a non-timesliced design for all CUs and the network.
We estimate area as that of Capstan~\cite{rucker2021capstan} with the logic from Aurochs~\cite{vilim2020gorgon} added in, giving a total area of approximately \SI{189}{mm^2} in a \SI{15}{nm} educational process with a clock frequency of \SI{1.6}{GHz}. 
Because the \SI{15}{nm} library lacks a memory compiler, prior work used SRAMs scaled from a \SI{28}{nm} industrial library.
This is 4.3\texttimes{} smaller than Nvidia's V100 GPU, our primary baseline~\cite{v100datasheet}.

To ensure that we accurately model hardware, we split \name{}'s compiled programs to map to the blocks provided by our vRDA machine model.
Our splitting constraints are the number of pipeline stages, registers, inputs, and outputs; we assume that merge units, constant inputs to merges, counters, and void inputs do not consume resources beyond their associated input buffers and registers.
Furthermore, to respect MU and AG mapping limits, we only map address generation contexts where all inputs are scalar and output-accumulation contexts where the only operation is a void reduction.
We further assume that operations can read and write 8- or 16-bit sub-registers 
and a small skid-buffer when reshaping vectors to fit on scalar links. 

\paragraph{Baselines}
We evaluate \name{} against an Nvidia V100~\cite{jia2018dissecting} GPU and an Intel Xeon CPU.
All of our applications have independent threads running under a parallel region, so we scale problem sizes across platforms so that each reaches its peak performance and report normalized performance in GB/s.
This ensures that baselines have the best performance possible and sets a lower bound on \name's performance. Application sizes are reported as the sum of input and output data sizes, except for kD-tree, which uses the size of the fetched points that are counted.

GPU tests were performed on an AWS p3.2xlarge instance using CUDA 11.6, RAPIDS 22.04~\cite{rapids}, and cuCollections~\cite{cuco} running Linux 5.13.0 and Nvidia driver 510.47.03. For all benchmarks except kD-tree, we use nvprof to measure only kernel runtime, which excludes device/host transfers, barriers, and CUDA stream synchronization. kD-tree uses host timers because the RAPIDS implementation uses multiple kernels. CPU tests were performed on an AWS m6i.16xlarge using GCC 11.2.0 with -O3 and OpenMP. The CPU is a \nth{3} generation Xeon Platinum at \SI{3.5}{GHz} with 64 threads and \SI{205}{GB/s} of DDR4 bandwidth. For both baselines, benchmarks were run 25 times and the average of all runs was taken. 

\paragraph{Applications}
We use a variety of applications to evaluate \name, as shown in \Cref{tab:apps_revet}.
These applications are selected to focus on \name{}'s \emph{new} functionality, so they all represent applications that cannot be compiled to Plasticine or other vRDAs. 
Because threads provide a superset of MapReduce functionality, any code that could be compiled by Spatial could also be mapped by \name.
They are drawn from a variety of domains including string analytics, data-structure traversal, search, and generic data-processing algorithms like hashing.
We focus on discrete kernels to avoid inter-kernel overheads in the baselines; the GPU tree traversal is the only multi-kernel baseline.

When evaluating applications, we assume that runtime is a function of bulk throughput, data size, and initialization time:
$\mathrm{runtime} = \mathrm{size}/\mathrm{throughput} + \mathrm{init}.$
Furthermore, all of our benchmarks exploit abundant, non-communicating threaded parallelism, so the amount of work can be increased without changing the \emph{nature} of the work.
\name{} also uses static SRAM allocation instead of caches, which means that threads do not interfere with each other, and adding more threads will not decrease aggregate system throughput.

Therefore, for every platform, we use the largest data sizes feasible to measure throughput, which yields trivially short initialization times.
Although the data sizes for \name{} are relatively small, the inclusion of initialization time means that throughput would only \emph{increase} with larger datasets.

\subsection{Resource Requirements \& Performance}
\name{} generates resource-efficient vRDA configurations.
Furthermore, using \name{}, a vRDA out-performs a GPU on a variety of applications and is DRAM-bandwidth-limited for many of them as well.
\paragraph{Resource Breakdown}
Our first evaluation shows the vRDA resources required by \name-generated code in \Cref{tab:resources}.
It is challenging for a vRDA application to make 100\% use of resources due to on-chip network constraints~\cite{zhang2019scalable}, so we scale outer parallelism to use 70\% usage of the critical resource (CU, MU, or AG).

Using outer and vector parallelism, \name{} can provide hundreds of SIMD-parallel lanes.
Furthermore, some applications (isipv4 and ip2int) are outer-parallelized at two levels: tile loads/stores for thread arguments/results and the inner loops.
For these, up to three vectorized streams (48 lanes) process tiles of thread inputs/outputs while thirty vectorized (480 lanes) streams process the inner while loops.
Although \name{} maps fewer vector lanes than a GPU has CUDA cores, \name{} lanes process multiple pipelined instructions per cycle.
\begin{table}
  \scriptsize
  \centering
  \setlength{\tabcolsep}{5pt}
  \caption{Performance evaluation, including tests with ideal models for SRAM (S), Network (N), and DRAM (D).}\label{tab:perf_revet}
  \begin{tabu}{Xrrrrrrrrrrr}
    \toprule
    & \name{} &  \multicolumn{2}{c}{V100} & \multicolumn{2}{c}{CPU} & \multicolumn{3}{c}{Ideal (Speedup \texttimes)} \\\cmidrule(lr){2-2}\cmidrule(lr){3-4}\cmidrule(lr){5-6}\cmidrule(l){7-9}
                   & GB/s & GB/s  & \texttimes & GB/s & \texttimes & D    & SN & SND \\
    \midrule                               
    isipv4    &    443&121&3.65&7.3&60.6&1.04&1.07&1.18\\
    ip2int    &508&381&1.33&9.1&55.9&1.42&1.03&1.55    \\
    murmur3   &628&218&2.88&122.2&5.1&1.55&1.07&2.37   \\
    hash-table&42&40&1.05&7.4&5.7&2.70&1.00&3.23       \\
    search    &481&51&9.45&120.6&4.0&1.37&1.18&1.38    \\
    huff-dec  &380&97&3.91&19.0&20.1&0.98&1.07&1.08    \\
    huff-enc  &409&172&2.38&35.0&11.7&1.01&1.17&1.18   \\
    kD-tree   &52&1.5&33.93&3.4&15.3&1.28&0.92&1.65    \\
    \midrule   
    geomean   &&&3.81&&13.9&1.35&1.06&1.59   \\
    \bottomrule
  \end{tabu}
  \vskip-.3\baselineskip
\end{table}

\name{} has minimal resource overhead: for all of our applications, most mapped CUs are used for inner-loop operations, and only a few CUs and MUs are used for workload distribution and buffering live values around replicates.
MUs are also used for retiming, but these can frequently be shared with those inserted for deadlock avoidance. 

\paragraph{Throughput}
\Cref{tab:perf_revet} shows each design's throughput.
On average, \name{} is 3.8\texttimes{} faster than the GPU; when estimated die area is taken into account, this gap grows to over 16\texttimes.
Furthermore, isipv4, ip2int, and murmur3 use over 75\% of the peak HBM2 bandwidth; hash-table is limited by DRAM activations.
The greater geomean performance improvement from ideal DRAM (+35\%, D) than ideal on-chip resources (+6\%, SN) also shows that our applications are well-mapped.

The GPU performs best when each thread processes a small amount of data (ip2int and isipv4 read about \SI{13}{B} per thread).
On applications like murmur3 and search, the GPU is slowed down by the longer data involved (\SI{64}{B} and \SI{256}{B}). 
Although SIMT supports 32 threads per cycle, the GPU can process fewer independent accesses through its L1 cache: therefore, independent threads cannot run at full throughput unless they access nearby addresses.
This is because the GPU expects, and requires, coalescing for cached levels of the memory hierarchy (i.e., everything except explicit SRAM): the L1 cache can only execute a certain number of tag checks per cycle~\cite{lloyd2019gpucheck}.
\name{} does not have this problem: because iterators are mapped in SRAM, they execute in parallel without tag checks.

Search performs better on the vRDA because \name{}'s support for efficient branching enables the asymptotically-efficient Boyer-Moore~\cite{boyer1977fast} algorithm.
Boyer-Moore is complex because each thread is independently matching backwards along the pattern or computing an offset; \name{} uses nested \lstinline!while! loops to support this behavior.
In addition to a poor search algorithm, the GPU's constraints also force a poor algorithm for tree traversal.
Because CUDA does not support recursion (like the CPU) or \lstinline!fork! statements (like \name), every iteration of its quad-tree traversal must write into a large array.
However, because each iteration only selects a few children, little parallelism is extracted to amortize inter-kernel overheads.

\paragraph{Aurochs Comparison}
Finally, we compare to Aurochs~\cite{vilim2021aurochs}, a primitive implementation of dataflow threads.
Most \name{} applications cannot run on Aurochs because it lacks support for the local allocation needed for intra-thread locality. The tree traversal benchmark is supported.
The more efficient on-chip primitives supported by \name{} allow the kD-tree implementation to be over 11\texttimes{} faster than the Aurochs tree traversal.
First, Aurochs lacked support for thread-local storage, which results in up to 10 live variables traversing its pipeline that have to be duplicated whenever threads are forked; \name{} can store these variables in SRAM.

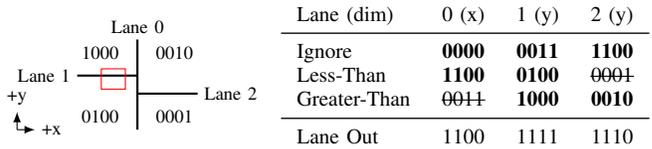
\begin{figure}
\begin{minipage}{0.4\linewidth}
  \centering
\scriptsize
\begin{tikzpicture}[y=0.6cm,x=0.8cm]
\draw[thick] (1,2) -- (1,0);
\draw[thick] (0,1.2) -- (1,1.2);
\draw[thick] (1,0.8) -- (2,0.8);
\node[anchor=south] at (1,2) {Lane 0};
\node[anchor=east] at (0,1.2) {Lane 1};
\node[anchor=west] at (2,0.8) {Lane 2};
\node[anchor=east] at (0.8,1.7) {1000};
\node[anchor=east] at (0.8,0.3) {0100};
\node[anchor=west] at (1.2,1.7) {0010};
\node[anchor=west] at (1.2,0.3) {0001};
\draw[->] (-1,0) -- (-1,0.4) node[above] {+y};
\draw[->] (-1,0) -- (-0.7,0) node[right] {+x};
\draw[thin, color=red] (0.4,1.35) -| (0.8, 0.9) -| cycle;
\end{tikzpicture}
\end{minipage}%
\begin{minipage}{0.6\linewidth}
  \centering
\footnotesize
\begin{tabular}{lccc}
Lane (dim) & 0 (x)  & 1 (y) & 2 (y) \\
\midrule
Ignore & \bfseries 0000 & \bfseries 0011 & \bfseries 1100 \\
Less-Than & \bfseries 1100 & \bfseries 0100 & \sout{0001} \\
Greater-Than & \sout{0011} & \bfseries 1000 & \bfseries 0010 \\
\midrule
Lane Out & 1100 & 1111 & 1110 \\
\end{tabular}
\end{minipage}
\caption{Three lanes work together to traverse a folded kD-tree. Each lane performs one comparison.}
\label{fig:kdtree}
\end{figure}

Second, Aurochs does not support fine-grained parallelism via \lstinline!foreach! loops.
Our kD-tree uses a \lstinline!foreach! loop to vectorize 15 comparisons for every node, which are ANDed together to identify which child nodes should be traversed.
\Cref{fig:kdtree} shows a simplified version that uses three lanes to traverse two tree levels.
Every lane's comparison starts with a mask for regions that are ignored: for instance, Lane~1 ignores the right two regions.
Then, the lane compares its partition value against the query's minimum and maximum ranges, which produces a per-lane output.
In the example, Lane~2's comparison, which produces a validity mask of 1110, is overridden by Lane~0, which determines that 0010 and 0001 are invalid.
Therefore, with only \SI{64}{B} loaded from DRAM per node, \name{} can handle a 16-ary tree.

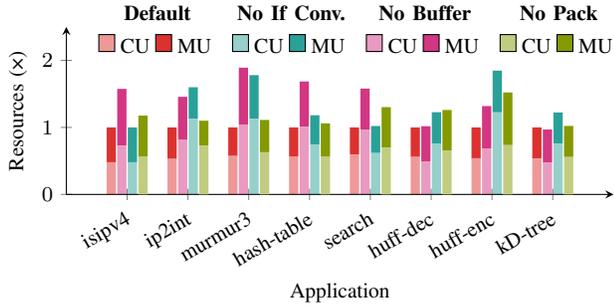
\begin{figure}
  \centering
  \begin{tikzpicture}
  \footnotesize
  \pgfplotsset{tmpstyle/.style={
    ybar stacked,
    bar width=3.5pt,
    height=1.5in,
    width=1.0\linewidth,
    ymin = 0,
    ymax = 2.5,
    xmin = -1,
    xmax = 8,
    axis lines = left,
    xlabel={Application},
    ylabel={Resources (\texttimes)},
    cycle list name=solarized-bars,
    xtick=data,
    xticklabels from table={data/opt_res.tsv}{App},
    xticklabel style={
      rotate=30,
      anchor=north east,
    },
    legend columns=2,
}}
  \begin{axis}[
    tmpstyle,
    bar shift=-6.0pt,
    hide axis,
    legend style={at={(axis cs:0.5,1.9)},above,name=legend,draw=none}
  ]
    \addplot table [x=XPos, y expr=(\thisrow{Opt-CU})/(\thisrow{Opt-CU}+\thisrow{Opt-MU}), col sep=space] {data/opt_res.tsv};
    \addplot table [x=XPos, y expr=(\thisrow{Opt-MU})/(\thisrow{Opt-CU}+\thisrow{Opt-MU}), col sep=space] {data/opt_res.tsv};
    \addlegendentry{CU}\addlegendentry{MU}
  \end{axis}
   \node [above,font=\bfseries] (legendtitle) at (legend.north) {Default};
  \begin{axis}[
    tmpstyle,
    bar shift=2.0pt,
    cycle list shift=2,
    hide axis,
    legend style={at={(axis cs:2.7,1.9)},above,name=legend,draw=none}
  ]
    \addplot table [x=XPos, y expr=(\thisrow{NoIfConv-CU})/(\thisrow{Opt-CU}+\thisrow{Opt-MU}), col sep=space] {data/opt_res.tsv};
    \addplot table [x=XPos, y expr=(\thisrow{NoIfConv-MU})/(\thisrow{Opt-CU}+\thisrow{Opt-MU}), col sep=space] {data/opt_res.tsv};
    \addlegendentry{CU}\addlegendentry{MU}
  \end{axis}
   \node [above,font=\bfseries] (legendtitle) at (legend.north) {No If Conv.};
  \begin{axis}[
    tmpstyle,
    bar shift=-2.0pt,
    cycle list shift=4,
    hide axis,
    legend style={at={(axis cs:4.9,1.9)},above,name=legend,draw=none}
  ]
    \addplot table [x=XPos, y expr=(\thisrow{NoBuf-CU})/(\thisrow{Opt-CU}+\thisrow{Opt-MU}), col sep=space] {data/opt_res.tsv};
    \addplot table [x=XPos, y expr=(\thisrow{NoBuf-MU})/(\thisrow{Opt-CU}+\thisrow{Opt-MU}), col sep=space] {data/opt_res.tsv};
    \addlegendentry{CU}\addlegendentry{MU}
  \end{axis}
   \node [above,font=\bfseries] (legendtitle) at (legend.north) {No Buffer};
  \begin{axis}[
    tmpstyle,
    bar shift=6.0pt,
    cycle list shift=6,
    legend style={at={(axis cs:7.1,1.9)},above,name=legend,draw=none}
  ]
    \addplot table [x=XPos, y expr=(\thisrow{NoPack-CU})/(\thisrow{Opt-CU}+\thisrow{Opt-MU}), col sep=space] {data/opt_res.tsv};
    \addplot table [x=XPos, y expr=(\thisrow{NoPack-MU})/(\thisrow{Opt-CU}+\thisrow{Opt-MU}), col sep=space] {data/opt_res.tsv};
    \addlegendentry{CU}\addlegendentry{MU}
  \end{axis}
   \node [above,font=\bfseries] (legendtitle) at (legend.north) {No Pack};
  \end{tikzpicture}
  \vskip-.6\baselineskip
  \caption{Resource increase (CUs and MUs) when turning off different \name{} optimization passes.}\label{fig:ressave}
\end{figure}
\subsection{Optimizations}
In \Cref{sec:implementation}, we discussed several optimization passes for \name.
In this section, we discuss how these either save resources or improve performance directly.

\paragraph{Resource-Saving Optimizations}
\Cref{fig:ressave} shows the effect of disabling enhanced if-to-select conversion, allocator hoisting and replicate bufferization, and variable packing.
Not all optimizations improve all applications: for example, \lstinline!if! to select conversion has no impact on isipv4, which has no convertible \lstinline!if! statements.
These passes lower resource requirements by either reducing the number of basic blocks (If Conv) or live variables that have to be permuted in the pipeline (Buffer and Pack).
However, extracting pointers after allocator hoisting can add resources, as seen in the Buffer column.
Without these resource-saving optimizations, only kD-tree would be able to hit the outer-parallelism factors that we target in our evaluation because it is AG-limited.

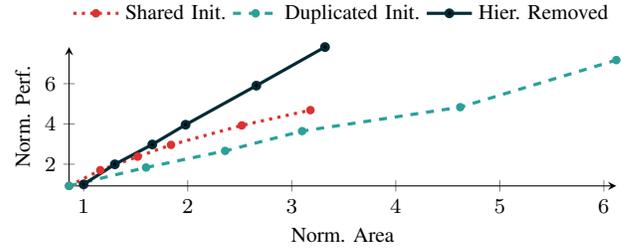
\begin{figure}
  \centering
  \begin{tikzpicture}
  \footnotesize
  \begin{axis}[
    axis lines = left,
    xlabel={Norm. Area},
    ylabel={Norm. Perf.},
    height=1.35in,
    ymin=0.922,
    xmin=0.86,
    width=\linewidth,
    cycle list name=solarized,
    legend style={at={(0.5,1.1)},anchor=south,draw=none},
    legend columns=3
  ]

  \addplot table [x=noflat-inner-area, y=noflat-inner-perf, col sep=comma] {data/elim_hier.csv};
  \addplot table [x=noflat-outer-area, y=noflat-outer-perf, col sep=comma] {data/elim_hier.csv};
  \addplot table [x=default-area, y=default-perf, col sep=comma] {data/elim_hier.csv};
  \addlegendentry{Shared Init.};
  \addlegendentry{Duplicated Init.};
    \addlegendentry{Hier. Removed};
  \end{axis}
  \end{tikzpicture}

  \caption{Performance vs. area with and without hierarchy removal (ideal SRAM, network, and DRAM models). Hierarchy removal moves the scaling curve up and left, with more performance at lower area.}\label{fig:hierremoval}
\end{figure}

\paragraph{Hierarchy Removal}
\Cref{fig:hierremoval} shows how hierarchy removal improves area-performance scaling, using murmur3 as a case study. 
Generally, every application loads a tile of thread initialization data from DRAM, computes the thread results, and stores the data back.
With hierarchy removal, applications run on small tiles which can coexist in the pipeline.
The hierarchical case uses large tiles that \emph{cannot} coexist: one tile must be done inside a while loop before another can start executing.
These large tiles can either be loaded outside the \lstinline!replicate! (shared initialization logic) or inside the \lstinline!replicate! (duplicated init.).

With tiles loaded outside replicated regions, hierarchy actually slightly reduces area by limiting overhead.
However, as outer-parallelism increases, the parallelism allocated to each replicated region decreases, which  leads to a widening performance gap.
If tiles are instead loaded inside replicated regions, hierarchy can achieve similar performance but with area increases from duplicated tile loads and stores.

\begin{figure}
  \centering
  \begin{tikzpicture}
  \footnotesize
  \begin{axis}[
    axis lines = left,
    xlabel={Input Elements},
    ylabel={Region Load (\%)},
    height=1.35in,
    width=1.0\linewidth,
    cycle list name=solarized-single,
    xmin=10000,
    xmax=9000000,
    ymin=9,
    ymax=16,
    ytick={10.0,12.5,15.0},
    yticklabels={10.0,12.5,15.0},
    xmode=log
  ]

  \addplot table [x=count, y expr=100*\thisrow{par0}/\thisrow{count}, col sep=space] {data/buf.tsv};
  \addplot table [x=count, y expr=100*\thisrow{par1}/\thisrow{count}, col sep=space] {data/buf.tsv};
  \addplot table [x=count, y expr=100*\thisrow{par2}/\thisrow{count}, col sep=space] {data/buf.tsv};
  \addplot table [x=count, y expr=100*\thisrow{par3}/\thisrow{count}, col sep=space] {data/buf.tsv};
  \addplot table [x=count, y expr=100*\thisrow{par4}/\thisrow{count}, col sep=space] {data/buf.tsv};
  \addplot table [x=count, y expr=100*\thisrow{par5}/\thisrow{count}, col sep=space] {data/buf.tsv};
  \addplot table [x=count, y expr=100*\thisrow{par6}/\thisrow{count}, col sep=space] {data/buf.tsv};
  \addplot table [x=count, y expr=100*\thisrow{par7}/\thisrow{count}, col sep=space] {data/buf.tsv};
    \node[anchor=west] (slow) at (axis cs:1100000,10) {slow (less work)};
    \node[anchor=west] (fast) at (axis cs:1100000,14.1) {fast (more work)};
    \draw[thick, <->] (slow) -- (fast);
  \end{axis}
  \end{tikzpicture}
  \lstset{style=revetfootnote}
  \caption{Per-region load vs. number of inputs for search. \lstinline!replicate! regions are assigned work based on throughput.}\label{fig:loadbalance}
\end{figure}
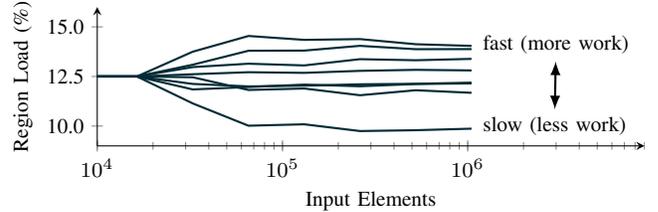

\paragraph{Load Balancing}
\Cref{fig:loadbalance} shows how allocation hoisting improves search's performance on the hybrid network, where different outer-parallel regions run at different throughputs.
This handles a slow \lstinline!replicate! region that is 30\% slower than the fastest one, and is better than Plasticine's~\cite{prabhakar2017plasticine} fixed work allocation, which would be bottlenecked by the slow region.
For small amounts of work (left), the allocator is able to assign buffers to all incoming threads without starving.
Therefore, every region gets an equal amount of work (12.5\% of the input).
As the amount of work increases, the allocator runs out of buffers, so not every thread is able to run in the first wave.
Then, faster regions free their allocated buffers first, and regions are assigned new work only after they complete existing work.
This creates a feedback loop that leads to slower regions receiving less work (less than 10\%) and faster regions more (14\%), avoiding a 21\% slowdown if all regions ran at the slowest region's speed.

\section{Related Work}
\label{sec:related}
In this section, we discuss how \name{} differs from prior work, including SIMT, programmable dataflow machines, scratchpad management, and parallel languages.

\paragraph{SIMT \& Vector-Threads}
SIMT models like CUDA~\cite{nvidia2013cuda} and Vector-Threads~\cite{krashinsky2004vector} are the dominant programming model for GPUs and the inspiration for \name{}.
In CUDA GPUs, 32 threads form a warp which can execute one instruction per cycle, so inactive threads waste an execution slot.
\name{} can avoid this problem using spatial execution.
Furthermore, CUDA's thread blocks prevent efficient dynamic thread spawning, while dataflow threads permits easy thread duplication in spatial pipelines~\cite{cudynamicparallelism}.

\paragraph{Programmable Dataflow}
Plasticine~\cite{prabhakar2017plasticine}, targeted by Spatial~\cite{koeplinger2018spatial} and SARA~\cite{zhang2021sara}, only has one FSM per compute unit. 
Therefore, one iteration has to complete before the next one can start!
HLS~\cite{coussy2009introduction} suffers from the same global-FSM problem, because it slices C programs into control logic (FSM) and datapath.
Aurochs~\cite{vilim2021aurochs} was the first vRDA to support dataflow threading using the relational-algebra operations introduced by Gorgon~\cite{vilim2020gorgon}.
Aurochs lacked composable control-flow primitives and as a result did not support high-level compilation.
Furthermore, Aurochs did not support per-thread SRAM buffers and could not send on-chip scalar values, both of which are needed for efficient dataflow.
Unlike Capstan~\cite{rucker2021capstan} and SAM~\cite{hsu2023sparse}, which enable direct loops over sparse data structures, \name{} optimizes for easier-to-achieve parallelism across threads.

Stream-join~\cite{nowatzki2017stream,weng2020dsagen}, as proposed in the SPU~\cite{dadu2019towards}, is another paradigm for dataflow computing, but also suffers from the single-FSM problem. 
Instruction-based designs (where a CGRA is integrated with a CPU, like SNAFU~\cite{gobieski2021snafu} and MANIC~\cite{gobieski2019manic}) suffer from the single-FSM problem as well, because the CPU can only logically execute one thread at a time.
Similarly, time-scheduling (e.g., Fifer~\cite{nguyen2021fifer}) virtualizes hardware to provide the abstraction of more resources without changing the underlying compute model.
Virtualizing multiple computing contexts can provide the abstraction of multiple FSMs, but the need for reconfiguration means that only one FSM can run at a time.

Other approaches like Fleet~\cite{thomas2020fleet} and CoRAM~\cite{chung2011coram} help support streaming on FPGAs but require RTL for the streaming algorithms.
Others have mapped complicated programs to tagged dataflow~\cite{arvind1977indeterminacy}, including Kahnian networks~\cite{khan1974semantics} and the Monsoon processor~\cite{papadopoulos1990monsoon}. 
\name{} is more efficient and powerful because it targets vectorized, pipelined dataflow without the need for tags and does so from an imperative language.

\paragraph{Data Orchestration}
By targeting dataflow hardware with scratchpads instead of caches, \name{} can improve efficiency relative to von Neumann approaches (albeit by eliminating the potential for reuse across threads).
Other approaches like Buffets~\cite{pellauer2019buffets} and Stash~\cite{komuravelli2015stash} also automatically orchestrate scratchpad memory hierarchies.
However, \name's approach is unique by natively supporting multithreaded accesses and reusing the logic of a vRDA to do so.

\paragraph{Languages \& Compilers}
Others have proposed streaming-native domain-specific languages (DSLs) to capture dataflow behavior like StreamIt~\cite{thies2002streamit} and Spidle~\cite{consel2003spidle}.
However, the goal of \name{} is not to expose streaming to the user, but rather to expose an imperative language and lower it to a dataflow backend.
Finally, Cilk~\cite{blumofe1996cilk} and OpenMP~\cite{dagum1998openmp} provide C extensions for parallelism, like \name{}; however, these languages use their extensions to target multicore CPUs and cannot lower to dataflow.

\section{Conclusion}
\label{sec:conclusion}
We introduce \name{}, a compiler that takes threaded imperative code and lowers it to run on a vectorized RDA.
\name{} enables control flow in the presence of abundant unordered parallelism on an architectural paradigm that previously only supported control-flow-free parallel sections.
Thus, \name{} provides SIMT's \emph{threaded} abstraction---one control flow decision per lane, per cycle---while also demonstrating intelligent scratchpad allocation that eliminates the need for caches for a wide range of threaded applications.
On a variety of real-world applications, \name{} is 3.8\texttimes{} faster than a GPU on a 4.3\texttimes{} smaller chip and over 13\texttimes{} faster than a CPU.

\section*{Acknowledgements}
We would like to thank Muhammad Shahbaz, Olivia Hsu, Tian Zhao, and the anonymous reviewers for their feedback on this paper.
This work was supported in part by the NSF under grant numbers 1937301, 2028602, CCF-1563078, and 1563113.  
This research was also supported in part by the Stanford Data Analytics for What’s Next (DAWN) Affiliate Program. Any opinions,
findings, and conclusions or recommendations expressed in this material are those of the authors
and do not necessarily reflect the views of the aforementioned funding agencies.

\bibliographystyle{IEEEtranS}
\bibliography{main}

\begin{thebibliography}{10}
\providecommand{\url}[1]{#1}
\csname url@samestyle\endcsname
\providecommand{\newblock}{\relax}
\providecommand{\bibinfo}[2]{#2}
\providecommand{\BIBentrySTDinterwordspacing}{\spaceskip=0pt\relax}
\providecommand{\BIBentryALTinterwordstretchfactor}{4}
\providecommand{\BIBentryALTinterwordspacing}{\spaceskip=\fontdimen2\font plus
\BIBentryALTinterwordstretchfactor\fontdimen3\font minus
  \fontdimen4\font\relax}
\providecommand{\BIBforeignlanguage}[2]{{%
\expandafter\ifx\csname l@#1\endcsname\relax
\typeout{** WARNING: IEEEtranS.bst: No hyphenation pattern has been}%
\typeout{** loaded for the language `#1'. Using the pattern for}%
\typeout{** the default language instead.}%
\else
\language=\csname l@#1\endcsname
\fi
#2}}
\providecommand{\BIBdecl}{\relax}
\BIBdecl

\bibitem{cudynamicparallelism}
\BIBentryALTinterwordspacing
A.~Adinets. {CUDA} dynamic parallelism {API} and principles. [Online].
  Available:
  \url{https://developer.nvidia.com/blog/cuda-dynamic-parallelism-api-principles/}
\BIBentrySTDinterwordspacing

\bibitem{arvind1977indeterminacy}
Arvind, K.~P. Gostelow, and W.~Plouffe, ``Indeterminacy, monitors, and
  dataflow,'' \emph{ACM SIGOPS Operating Systems Review}, vol.~11, no.~5, pp.
  159--169, 1977.

\bibitem{blumofe1996cilk}
R.~D. Blumofe, C.~F. Joerg, B.~C. Kuszmaul, C.~E. Leiserson, K.~H. Randall, and
  Y.~Zhou, ``Cilk: An efficient multithreaded runtime system,'' \emph{Journal
  of Parallel and Distributed Computing}, vol.~37, no.~1, pp. 55--69, 1996.

\bibitem{boyer1977fast}
R.~S. Boyer and J.~S. Moore, ``A fast string searching algorithm,''
  \emph{Communications of the ACM}, vol.~20, no.~10, pp. 762--772, 1977.

\bibitem{buck2004brook}
I.~Buck, T.~Foley, D.~Horn, J.~Sugerman, K.~Fatahalian, M.~Houston, and
  P.~Hanrahan, ``Brook for {GPUs:} stream computing on graphics hardware,''
  \emph{ACM Transactions on Graphics (TOG)}, vol.~23, no.~3, pp. 777--786,
  2004.

\bibitem{chung2011coram}
E.~S. Chung, J.~C. Hoe, and K.~Mai, ``{CoRAM:} an in-fabric memory architecture
  for {FPGA-based} computing,'' in \emph{Proceedings of the 19th ACM/SIGDA
  International Symposium on Field Programmable Gate Arrays}, 2011, pp.
  97--106.

\bibitem{consel2003spidle}
C.~Consel, H.~Hamdi, L.~R{\'e}veill{\`e}re, L.~Singaravelu, H.~Yu, and C.~Pu,
  ``Spidle: A {DSL} approach to specifying streaming applications,'' in
  \emph{International Conference on Generative Programming and Component
  Engineering}.\hskip 1em plus 0.5em minus 0.4em\relax Springer, 2003, pp.
  1--17.

\bibitem{coussy2009introduction}
P.~Coussy, D.~D. Gajski, M.~Meredith, and A.~Takach, ``An introduction to
  high-level synthesis,'' \emph{IEEE Design \& Test of Computers}, vol.~26,
  no.~4, pp. 8--17, 2009.

\bibitem{dadu2019towards}
V.~Dadu, J.~Weng, S.~Liu, and T.~Nowatzki, ``Towards general purpose
  acceleration by exploiting common data-dependence forms,'' in
  \emph{Proceedings of the 52nd Annual IEEE/ACM International Symposium on
  Microarchitecture}, 2019, pp. 924--939.

\bibitem{dagum1998openmp}
L.~Dagum and R.~Menon, ``{OpenMP}: an industry standard {API} for shared-memory
  programming,'' \emph{IEEE computational science and engineering}, vol.~5,
  no.~1, pp. 46--55, 1998.

\bibitem{gobieski2021snafu}
G.~Gobieski, A.~O. Atli, K.~Mai, B.~Lucia, and N.~Beckmann, ``Snafu: an
  ultra-low-power, energy-minimal {CGRA-generation} framework and
  architecture,'' in \emph{2021 ACM/IEEE 48th Annual International Symposium on
  Computer Architecture (ISCA)}.\hskip 1em plus 0.5em minus 0.4em\relax IEEE,
  2021, pp. 1027--1040.

\bibitem{gobieski2019manic}
G.~Gobieski, A.~Nagi, N.~Serafin, M.~M. Isgenc, N.~Beckmann, and B.~Lucia,
  ``Manic: A vector-dataflow architecture for ultra-low-power embedded
  systems,'' in \emph{Proceedings of the 52nd Annual IEEE/ACM International
  Symposium on Microarchitecture}, 2019, pp. 670--684.

\bibitem{hsu2023sparse}
O.~Hsu, M.~Strange, R.~Sharma, J.~Won, K.~Olukotun, J.~S. Emer, M.~A. Horowitz,
  and F.~Kj{\o}lstad, ``The sparse abstract machine,'' in \emph{Proceedings of
  the 28th ACM International Conference on Architectural Support for
  Programming Languages and Operating Systems, Volume 3}, 2023, pp. 710--726.

\bibitem{standard2013high}
{JEDEC}, ``High bandwidth memory {(HBM) DRAM},'' \emph{Jesd235}, vol.~16, 2013.

\bibitem{jia2018dissecting}
Z.~Jia, M.~Maggioni, B.~Staiger, and D.~P. Scarpazza, ``Dissecting the {NVIDIA
  Volta GPU} architecture via microbenchmarking,'' \emph{arXiv preprint
  arXiv:1804.06826}, 2018.

\bibitem{jouppi2021ten}
N.~P. Jouppi, D.~H. Yoon, M.~Ashcraft, M.~Gottscho, T.~B. Jablin, G.~Kurian,
  J.~Laudon, S.~Li, P.~Ma, X.~Ma \emph{et~al.}, ``Ten lessons from three
  generations shaped {Google's TPUv4i:} industrial product,'' in \emph{2021
  ACM/IEEE 48th Annual International Symposium on Computer Architecture
  (ISCA)}.\hskip 1em plus 0.5em minus 0.4em\relax IEEE, 2021, pp. 1--14.

\bibitem{jouppi2017datacenter}
N.~P. Jouppi, C.~Young, N.~Patil, D.~Patterson, G.~Agrawal, R.~Bajwa, S.~Bates,
  S.~Bhatia, N.~Boden, A.~Borchers \emph{et~al.}, ``In-datacenter performance
  analysis of a tensor processing unit,'' in \emph{Proceedings of the 44th
  Annual International Symposium on Computer Architecture}, 2017, pp. 1--12.

\bibitem{khan1974semantics}
G.~Khan, ``The semantics of a simple language for parallel programming,''
  \emph{Information Processing}, vol.~74, pp. 471--475, 1974.

\bibitem{kim2015ramulator}
Y.~Kim, W.~Yang, and O.~Mutlu, ``Ramulator: A fast and extensible {DRAM}
  simulator,'' \emph{IEEE Computer Architecture Letters}, vol.~15, no.~1, pp.
  45--49, 2015.

\bibitem{koeplinger2018spatial}
D.~Koeplinger, M.~Feldman, R.~Prabhakar, Y.~Zhang, S.~Hadjis, R.~Fiszel,
  T.~Zhao, L.~Nardi, A.~Pedram, C.~Kozyrakis \emph{et~al.}, ``Spatial: A
  language and compiler for application accelerators,'' in \emph{Proceedings of
  the 39th ACM SIGPLAN Conference on Programming Language Design and
  Implementation}, 2018, pp. 296--311.

\bibitem{komuravelli2015stash}
R.~Komuravelli, M.~D. Sinclair, J.~Alsop, M.~Huzaifa, M.~Kotsifakou,
  P.~Srivastava, S.~V. Adve, and V.~S. Adve, ``Stash: Have your scratchpad and
  cache it too,'' \emph{ACM SIGARCH Computer Architecture News}, vol.~43,
  no.~3S, pp. 707--719, 2015.

\bibitem{krashinsky2004vector}
R.~Krashinsky, C.~Batten, M.~Hampton, S.~Gerding, B.~Pharris, J.~Casper, and
  K.~Asanovic, ``The vector-thread architecture,'' in \emph{Proceedings. 31st
  Annual International Symposium on Computer Architecture, 2004.}\hskip 1em
  plus 0.5em minus 0.4em\relax IEEE, 2004, pp. 52--63.

\bibitem{lattner2020mlir}
C.~Lattner, M.~Amini, U.~Bondhugula, A.~Cohen, A.~Davis, J.~Pienaar, R.~Riddle,
  T.~Shpeisman, N.~Vasilache, and O.~Zinenko, ``{MLIR:} a compiler
  infrastructure for the end of {Moore's} law,'' \emph{arXiv preprint
  arXiv:2002.11054}, 2020.

\bibitem{lloyd2019gpucheck}
T.~Lloyd, K.~Ali, and J.~N. Amaral, ``{GPUCheck:} detecting {CUDA} thread
  divergence with static analysis,'' 2019.

\bibitem{nguyen2021fifer}
Q.~M. Nguyen and D.~Sanchez, ``Fifer: Practical acceleration of irregular
  applications on reconfigurable architectures,'' in \emph{MICRO-54: 54th
  Annual IEEE/ACM International Symposium on Microarchitecture}, 2021, pp.
  1064--1077.

\bibitem{nowatzki2017stream}
T.~Nowatzki, V.~Gangadhar, N.~Ardalani, and K.~Sankaralingam, ``Stream-dataflow
  acceleration,'' in \emph{2017 ACM/IEEE 44th Annual International Symposium on
  Computer Architecture (ISCA)}.\hskip 1em plus 0.5em minus 0.4em\relax IEEE,
  2017, pp. 416--429.

\bibitem{cuco}
\BIBentryALTinterwordspacing
Nvidia. {cuCollections}. [Online]. Available:
  \url{https://github.com/NVIDIA/cuCollections}
\BIBentrySTDinterwordspacing

\bibitem{v100datasheet}
\BIBentryALTinterwordspacing
------. Nvidia {Tesla V100 GPU} architecture. [Online]. Available:
  \url{https://images.nvidia.com/content/volta-architecture/pdf/volta-architecture-whitepaper.pdf}
\BIBentrySTDinterwordspacing

\bibitem{rapids}
\BIBentryALTinterwordspacing
------. Rapidsai. [Online]. Available: \url{https://docs.rapids.ai/}
\BIBentrySTDinterwordspacing

\bibitem{nvidia2013cuda}
------, ``{CUDA C} programming guide,'' 2013.

\bibitem{papadopoulos1990monsoon}
G.~M. Papadopoulos and D.~E. Culler, ``Monsoon: an explicit token-store
  architecture,'' \emph{ACM SIGARCH Computer Architecture News}, vol.~18, no.
  2SI, pp. 82--91, 1990.

\bibitem{parr2013definitive}
T.~Parr, \emph{The definitive {ANTLR} 4 reference}.\hskip 1em plus 0.5em minus
  0.4em\relax Pragmatic Bookshelf, 2013.

\bibitem{pellauer2019buffets}
M.~Pellauer, Y.~S. Shao, J.~Clemons, N.~Crago, K.~Hegde, R.~Venkatesan, S.~W.
  Keckler, C.~W. Fletcher, and J.~Emer, ``Buffets: An efficient and composable
  storage idiom for explicit decoupled data orchestration,'' in
  \emph{Proceedings of the Twenty-Fourth International Conference on
  Architectural Support for Programming Languages and Operating Systems}, 2019,
  pp. 137--151.

\bibitem{prabhakar2021sambanova}
R.~Prabhakar and S.~Jairath, ``{SambaNova SN10 RDU:} accelerating software 2.0
  with dataflow,'' in \emph{2021 IEEE Hot Chips 33 Symposium (HCS)}.\hskip 1em
  plus 0.5em minus 0.4em\relax IEEE, 2021, pp. 1--37.

\bibitem{prabhakar2017plasticine}
R.~Prabhakar, Y.~Zhang, D.~Koeplinger, M.~Feldman, T.~Zhao, S.~Hadjis,
  A.~Pedram, C.~Kozyrakis, and K.~Olukotun, ``Plasticine: A reconfigurable
  architecture for parallel patterns,'' in \emph{2017 ACM/IEEE 44th Annual
  International Symposium on Computer Architecture (ISCA)}.\hskip 1em plus
  0.5em minus 0.4em\relax IEEE, 2017, pp. 389--402.

\bibitem{rucker2021capstan}
A.~Rucker, M.~Vilim, T.~Zhao, Y.~Zhang, R.~Prabhakar, and K.~Olukotun,
  ``Capstan: A vector {RDA} for sparsity,'' in \emph{MICRO-54: 54th Annual
  IEEE/ACM International Symposium on Microarchitecture}, 2021, pp. 1022--1035.

\bibitem{schardl2017tapir}
T.~B. Schardl, W.~S. Moses, and C.~E. Leiserson, ``Tapir: Embedding fork-join
  parallelism into {LLVM's} intermediate representation,'' in \emph{Proceedings
  of the 22Nd ACM SIGPLAN Symposium on Principles and Practice of Parallel
  Programming}, 2017, pp. 249--265.

\bibitem{posixfork}
\BIBentryALTinterwordspacing
{The Open Group}. fork. [Online]. Available:
  \url{https://pubs.opengroup.org/onlinepubs/009696799/functions/fork.html}
\BIBentrySTDinterwordspacing

\bibitem{thies2002streamit}
W.~Thies, M.~Karczmarek, and S.~Amarasinghe, ``{StreamIt:} a language for
  streaming applications,'' in \emph{International Conference on Compiler
  Construction}.\hskip 1em plus 0.5em minus 0.4em\relax Springer, 2002, pp.
  179--196.

\bibitem{thomas2020fleet}
J.~Thomas, P.~Hanrahan, and M.~Zaharia, ``Fleet: A framework for massively
  parallel streaming on {FPGAs},'' in \emph{Proceedings of the Twenty-Fifth
  International Conference on Architectural Support for Programming Languages
  and Operating Systems}, 2020, pp. 639--651.

\bibitem{vilim2021aurochs}
M.~Vilim, A.~Rucker, and K.~Olukotun, ``Aurochs: An architecture for dataflow
  threads,'' in \emph{2021 ACM/IEEE 48th Annual International Symposium on
  Computer Architecture (ISCA)}.\hskip 1em plus 0.5em minus 0.4em\relax IEEE,
  2021, pp. 402--415.

\bibitem{vilim2020gorgon}
M.~Vilim, A.~Rucker, Y.~Zhang, S.~Liu, and K.~Olukotun, ``Gorgon: Accelerating
  machine learning from relational data,'' in \emph{2020 ACM/IEEE 47th Annual
  International Symposium on Computer Architecture (ISCA)}.\hskip 1em plus
  0.5em minus 0.4em\relax IEEE, 2020, pp. 309--321.

\bibitem{weng2020dsagen}
J.~Weng, S.~Liu, V.~Dadu, Z.~Wang, P.~Shah, and T.~Nowatzki, ``{DSAGEN:}
  synthesizing programmable spatial accelerators,'' in \emph{2020 ACM/IEEE 47th
  Annual International Symposium on Computer Architecture (ISCA)}.\hskip 1em
  plus 0.5em minus 0.4em\relax IEEE, 2020, pp. 268--281.

\bibitem{zhang2019scalable}
Y.~Zhang, A.~Rucker, M.~Vilim, R.~Prabhakar, W.~Hwang, and K.~Olukotun,
  ``Scalable interconnects for reconfigurable spatial architectures,'' in
  \emph{2019 ACM/IEEE 46th Annual International Symposium on Computer
  Architecture (ISCA)}.\hskip 1em plus 0.5em minus 0.4em\relax IEEE, 2019, pp.
  615--628.

\bibitem{zhang2021sara}
Y.~Zhang, N.~Zhang, T.~Zhao, M.~Vilim, M.~Shahbaz, and K.~Olukotun, ``{SARA:}
  scaling a reconfigurable dataflow accelerator,'' in \emph{2021 ACM/IEEE 48th
  Annual International Symposium on Computer Architecture (ISCA)}.\hskip 1em
  plus 0.5em minus 0.4em\relax IEEE, 2021, pp. 1041--1054.

\end{thebibliography}

\end{document}